\documentclass[12pt]{article}                   
\usepackage{latexsym,afterpage,epsfig,array,tabularx}
\def\thickone{\mbox{$1\!\!1$}}
\def\-{{\bf --}} 
\def\theequation{\arabic{section}.\arabic{equation}}
\def\V{|V\rangle}\def\W{\langle W|}
\def\Ewv#1{\W#1\V}\def\Ewvi{\langle W\V}
\def\systau{\tau}
\def\A{{\cal A}}\def\B{{\cal B}}\def\C{{\cal C}}
\def\E{{\cal E}}\def\F{{\cal F}}

\def\sp{{\rm sp}}\def\scp{{\rm scp}}
\def\Vs{|V_1\rangle}\def\Ws{\langle W_1|}
\def\rv#1{|#1\rangle}\def\lv#1{\langle#1|}
\def\(#1){(\ref{#1})}
\def\lhd{\lambda_{\rm hd}}\def\lld{\lambda_{\rm ld}}
\def\lmc{\lambda_{\rm mc}}
\begin{document}
\begin{center}
\quad\\
  {\LARGE Exact solution of a cellular automaton for traffic}
\quad \\ \quad \\ \quad \\
{\Large M.~R.~Evans,{\footnote {\tt e-mail: martin@ph.ed.ac.uk}} 
N.~Rajewsky,{\footnote {\tt e-mail: rajewsky@math.rutgers.edu}} and
E.~R.~Speer {\footnote {\tt e-mail: speer@math.rutgers.edu}}}
\end{center}
\vskip1cm 
\begin{center}
$^1$\, Department of Physics and Astronomy\\
University of Edinburgh, Mayfield Road\\ Edinburgh EH9 3JZ, U.K. \\
\quad \\
${^{2,}} {^3}$\,Department of Mathematics\\ Rutgers University,
New Brunswick\\ New Jersey 08903, USA.
\end{center}
\vskip0.3cm \begin{center} {\large {\bf ABSTRACT}}\end{center}
We present an exact solution of a probabilistic cellular
automaton for traffic with open boundary
conditions, e.g. cars can enter and leave a part of a highway
with certain probabilities. The model studied is the
asymmetric exclusion process (ASEP)
with {\it simultaneous} updating of all sites.  It is equivalent to
a special case ($v_{\rm max}=1$) of the
Nagel-Schreckenberg model for highway traffic, which has found many 
applications in real-time traffic simulations.
The simultaneous updating induces additional strong short range
correlations compared to other updating schemes.
The stationary state is written in terms of a matrix 
product solution. The corresponding algebra, which expresses a system-size
recursion relation for the weights of the configurations, is quartic,
in contrast to previous cases, in which
the algebra is quadratic. We derive the phase diagram and compute various
properties such as density profiles, two point functions 
and the fluctuations in the number of particles (cars) in the system.
The current and the density profiles can be mapped onto the ASEP
with other time discrete updating procedures. Through use of this
mapping, our results
also give new results for these models.
\section{Introduction}
\label{introduction}

 In this paper we study a simple probabilistic cellular automaton
which describes the flow---of particles, automobiles, or some other
conserved quantity---through a one dimensional system.  The particles
(or cars) of the model move on a finite lattice; at integer times they
{\it simultaneously} attempt to hop one site forward, succeeding with
probability $p$ if the site in front of them is empty.
We are interested in the case of open boundary conditions (OBC), in which,
simultaneously with the hopping of particles along the lattice, a particle
enters the system with probability $\alpha$ at the leftmost site if that
site is empty, and if the rightmost site is occupied then the particle on
that site exits with probability $\beta$.
We use a matrix product ansatz to give a complete solution
of this model.

In recent years, cellular automata models for traffic flow have gained much
attention, because they make real time traffic simulations possible (see
\cite{procs97} and references therein).  The model studied here is one such;
for example, if the injection probability $\alpha$ is large the model may be
regarded as describing the situation, familiar from everyday experience, of
the reduction of a two-lane to a one-lane road by, e.g., the presence of
construction work on one lane.  It is a special case of the well-known
Nagel-Schreckenberg \cite{traffic} model, obtained by requiring that the
parameter $v_{\rm max}$ of that model satisfy $v_{\rm max}=1$ so that cars
move at most one lattice spacing at each integer time.  In realistic computer
simulations of highway traffic, the Nagel-Schreckenberg model is usually used
with $v_{\rm max}=5$.  However, in the case of OBC the phase diagram and
density profiles are essentially independent of $v_{\rm max}$ \cite{santen},
and in general it has been observed that, for modeling city traffic, it is
sufficient to set $v_{\rm max}=1$ \cite{chopard}.

 The model is in fact a synchronous update version of the asymmetric
exclusion process (ASEP), widely studied in both the physics and
mathematics literature \cite{spitzer,liggett,spohnbook}.  The ASEP was
originally introduced as an interacting particle system evolving in
continuous time; this evolution is equivalent to the random sequential
update (RSU) procedure, in which randomly chosen particles hop one at
a time.  Other updating schemes have also been introduced, including
sublattice-parallel \cite{KDN,Schuetz} and ordered sequential
procedures \cite{work}, and the fully parallel updating (PU) scheme,
which corresponds to the probabilistic cellular automaton described
above.  See Section 9 and \cite{RSSS} for precise definitions and a
review of current knowledge about these models.  The ASEP with RSU and
OBC has been exactly solved \cite{DDM,DEHP,SD}, using (among other
methods) the matrix product ansatz, and these results have been
extended to the sublattice-parallel and ordered sequential updating
schemes \cite{hinri,work,honi,RSSS}.  The model with PU has proved to be less
tractable; for example, parallel updating can induce strong short 
range correlations \cite{schaschr} (these are absent under other 
updating schemes).  On the other hand, parallel
updating is important in practice for traffic modeling, both for
efficiency---the PU is usually much faster than the RSU---and for
effectiveness; for example, the Nagel-Schreckenberg model is always
implemented with PU, since that method has been found to give the best
agreement with measurements on freeway traffic \cite{traffic}.

We remark that if one writes $\beta = p \widetilde{\beta}$ and $\alpha = p
\widetilde{\alpha}$ and then takes the limit $p\to 0$, the model reduces to
the ASEP with random sequential updating and injection and extraction rates
$\widetilde{\beta}$ and $\widetilde{\alpha}$, respectively.

Let us now briefly discuss known results which are related to our
model.  For the parallel dynamics on a ring (that is, with periodic
boundary conditions), the exact solution was given in \cite{traffic};
here, in contrast to other updating schemes, the stationary state is
not a simple product measure---occupation numbers at distinct sites
are correlated.  For example, if $p$ is 1 and the density is $1/2$
then the stationary state consists of free flowing particle-hole
pairs, i.e., there is a strong particle-hole attraction.  For $p=1$,
the dynamics is equivalent to rule 184 for cellular automata, for
which transient properties have been analyzed \cite{Fuks}.  The steady
state for arbitrary $p$ and overall densities is obtained by
factorizing the weights of the configurations into clusters of length
two \cite{schaschr}; the strong short range correlations persist.  The
steady state for the generalization of the model where each particle
has its own hopping probability has also been solved \cite{evansneu}.
  
Tilstra and Ernst \cite{TE} studied the case of OBC and $p=1$. They
obtained results which they argue to be asymptotically (i.e., in the limit
of large system size) correct. In \cite{RSSS}, the system was found to be
exactly solvable on a special line in the phase diagram; from this special
case and extensive Monte Carlo simulations, the phase diagram and formulae
for the current and the bulk densities were conjectured.

We now discuss briefly the nature of our solution.  For the random
sequential model, the initial breakthrough was the observation that there
exists a recursion relation relating steady state weights (unnormalized
probabilities) for a system of size $N$ to those for a system of size $N-1$
\cite{DDM}. Equivalently, one may write the weights as matrix elements or
traces of products of operators; requiring these operators to satisfy
certain algebraic rules then implies that the weights satisfy the recursion
relations \cite{DEHP}. The matrix product allows a more direct
calculation of steady state correlation functions than the recursion
relations \cite{DDM,DE,SD}.

For the present model we have followed a similar line of attack.  When we
write the weights as operator products, however, we must require that the
operators satisfy quartic algebraic relations, which relate a product of
four operators to sums of products of three and two operators
(see section 2). This is in
contrast with the quadratic relations found in previous works
\cite{DEHP,DErev,evansetal,DJLS,mallick,DLS,sasa,rittenberg,ADR,Karim}.
For example, the recursion relation for the
weight $f_N(\ldots 0 1 0 0 \ldots)$ relates 
systems of size $N$ to
systems of size $N-1$ and $N-2$ in the following way:
\begin{equation}
f_N(\ldots 0 1 0 0 \ldots)
=(1-p) f_{N-1}(\ldots 0 1 0 \ldots) + f_{N-1}(\ldots 0 0 0 \ldots)
+pf_{N-2}(\ldots 0  0 \ldots)\quad.
\label{example_recur}
\end{equation}
We believe that the above method (recursion relations
in the system size) should be of general
interest as an analytic approach to probabilistic cellular automata
\cite{Rujan,GlD,LMS}, 
for which there are notoriously few exact results (see 
Schadschneider in \cite{procs97}).

We now summarize the content of the paper.  In
section~\ref{fundamental} we define the model and provide the
algebraic rules for a matrix product solution.  The next section gives
the proof that they indeed describe the stationary state; the argument
proceeds  by looking at
blocks of consecutive particles and holes.  In section~\ref{reduction}
we show that the quartic algebraic rules may be reduced to quadratic
rules by assuming the operators are two by two matrices whose elements
are matrices, generally of infinite dimension, i.e., that the
operators are rank four tensors,

The reduction to quadratic algebraic rules allows us to relate
the parallel update model to the model with other
discrete-time updating schemes.  Specifically in
section 5 we show that the current and density profile for parallel
update are simply related to those quantities for ordered sequential
and sublattice parallel updating, although the relation between higher
order correlation functions is more complicated.  Thus, in solving
exactly the parallel model in
sections~\ref{phasediagram}---\ref{combinatorial}, we also obtain new
exact results and prove conjectures for the other discrete-time models,
for which only the asymptotic current was previously known.

In section~\ref{representations} several explicit representations of
the matrices are constructed.  As a first application, we solve the
case $p=1$ by means of $4\times 4$ matrices in
section~\ref{pequalone}.  A detailed analysis is made of the two point
correlation functions in order to highlight the oscillating decay of
the correlation function which is a particular feature under parallel
dynamics.

We then turn to the task of obtaining the exact solution for general
$p<1$.  The current phase diagram is derived using generating function
techniques section~\ref{phasediagram}, and the asymptotic behavior of
the density profiles in section~\ref{asymptotics} again using
generating function techniques.  The relevant Tauberian theorems are
presented in appendix A.  For finite systems, we calculate exact
combinatorial expressions for the density profiles and two point
functions (section~\ref{combinatorial}). Technical details of the
computations are contained in appendix B. By combining the results
from the two preceding sections, we compute the asymptotic
bulk densities in section~\ref{bulk}. A discussion closes the
paper.
\setcounter{equation}{0}

 \section{Model Definition and Steady State Recursion Relations}
 \label{fundamental}

 In this paper we study the asymmetric exclusion process with parallel
dynamics and open boundary conditions.  We consider a one dimensional
lattice, with $N$ sites labelled $1$ through $N$. Each site $i$ may be
occupied by a particle, in which case a binary variable $\tau_i$ satisfies
$\tau_i=1$, or empty, in which case $\tau_i=0$. The n-tuple
$\systau=(\tau_1,\ldots,\tau_N)$ specifies the configuration of the system.
The dynamics is defined by requiring that at each time step three things
happen: (i)~{\em all} particles on sites $1,\ldots,N-1$ with an empty site
in front of them attempt to hop forwards, succeeding with probability $p$;
(ii)~if site 1 is empty then a particle attempts to enter the lattice
there, succeeding with probability $\alpha$; and (iii)~if site $N$ is
occupied then the particle there attempts to exit the lattice, succeeding
with probability $\beta$.  All of these processes are stochastically
independent. Note the particle-hole symmetry: the removal of a particle at
the right end can be viewed as an injection of a hole, so the dynamics is
invariant under the combined operations of interchange of $i$ and $N-i+1$,
interchange of particles and holes, and interchange of $\alpha$ and
$\beta$. For example, we have
\begin{equation}
\langle\tau_i\rangle_N(\alpha,\beta,p)
  =1-\langle\tau_{N+1-i}\rangle_N(\beta,\alpha,p)\,.
\label{symm}
\end{equation}

For $p$, $\alpha$, and $\beta$ nonzero, the configuration
$1\,0\,1\,0\,\ldots$ can be reached from any other, so the model, viewed as
a finite state Markov chain, has a single irreducible component and hence a
unique steady state \cite{Feller}, which we denote by $P_N$;
$P_N( \systau)$ is the probability of finding a system of size $N$ in
configuration $\systau$ in the long time limit. In calculating
$P_N( \systau)$ it is convenient, as noted in earlier work on the random
sequential model \cite{DDM,DE,DEHP,SD}, first to define unnormalized weights
$f_N( \systau)$ and then to recover the probabilities via
 \begin{equation}
    P_N(\systau) = f_N(\systau)/Z_N\, ,\label{PN} 
 \end{equation}
 where
 \begin{equation}
     Z_N = \sum_\tau f_N(\systau)\, ,\label{ZN}
 \end{equation}
 the sum taken over all configurations of size $N$.

The idea is now to  introduce a {\em matrix
product ansatz} by writing
\begin{equation}\label{weight}
    f_N(\systau)=\langle W|(\prod_{i=1}^{N}(1-\tau_i)E+\tau_iD)|V\rangle\;.
\end{equation}
 This is to be read as a product of operators $E$ and $D$ (an $E$ for each
empty site and a $D$ for each occupied site) contracted with a vector $\V$
and dual vector $\W$, yielding a scalar steady state weight;  for example,
if $N=6$ and $\systau=0\,1\,0\,0\,0\,1$ then $f_N(\systau)=\W EDEEED\V$.
This method originated in work on the random sequential model \cite{DEHP},
in which the operators were represented by infinite dimensional matrices.
Later papers generalizing the original idea have employed operators
represented by finite dimensional matrices \cite{essler,mallicksandow} or
higher rank infinite dimensional tensors \cite{DEM,EKKM}; both these
approaches will be used here.

  The operators $E$ and $D$ and vectors $\V$ and $\W$ are required to
satisfy certain algebraic rules, listed below.  We determined these
rules by finding the steady state explicitly for small system sizes
and then guessing.  We prove in the next section that these rules do
indeed imply that the steady state of the system is given by
(\ref{PN}) and (\ref{weight}) for general $N$, and in
section~\ref{representations} we
construct an explicit representation, thus verifying that the rules
are consistent.

 The rules for the bulk are
\begin{eqnarray}
{ EDEE} &=& (1-p){ EDE} + { EEE} + p{ EE}\, ,
\label{bulk1}\\
 { EDED} &=&  { EDD} + { EED} + p{ ED}\, ,
\label{bulk2}  \\
{ DDEE} &=& (1-p){ DDE} + (1-p){ DEE} + p(1-p){ DE}\, ,
 \label{bulk3}\\
{ DDED} &=& { DDD} + (1-p){ DED} + p{ DD}\, . 
\label{bulk4}
\end{eqnarray}
We also have rules involving three sites next to each boundary,
\begin{eqnarray}
{ DDE}|{ V} \rangle &=& (1-\beta){ DD}|{  V}\rangle
 + (1-p){ DE}| { V}\rangle + p(1-\beta){ D} | { V}\rangle\, , 
\label{DDEV} \\
{ EDE}|{ V} \rangle &=& (1-\beta){ ED}|{  V}\rangle
 + { EE}| { V}\rangle + p{ E} | { V}\rangle\, ,
\label{EDEV} \\
\langle W|DEE  &=& (1-\alpha)\langle W | EE
+(1-p)\langle W | DE  + p(1-\alpha)\langle W | E\, ,
\label{WDEE}\\
\langle W|DED  &=& (1-\alpha)\langle W | ED
+\langle W |DD  + p\langle W | D\, ,
\label{WDED}
\end{eqnarray}
and two  sites next to each boundary,
\begin{eqnarray}
{ DD} |{ V}\rangle &=& \frac{p(1-\beta)}{\beta} { D} |{ V}\rangle\, ,
\label{DDV}
\\
{ ED} |{ V}\rangle &=&  \frac{p}{\beta} { E} |{ V}\rangle \, ,
\label{EDV}
\\
\langle W| EE  &=& \frac{p(1-\alpha)}{\alpha}\langle W | E\, ,
\label{WEE}
\\
\langle W | ED  &=& \frac{p}{\alpha}\langle W| D\, .
\label{WED}
\end{eqnarray}
These rules permit the computation of all $f_N(\systau)$ (up to an overall
constant, which must be assumed to be nonzero). However,
a rather indirect argument, which we now discuss,
is required when $N=1$ or $N=2$.  It is convenient, and represents no loss of
generality, to assume that
 \begin{equation} 
 \Ewvi > 0. \label{WVne0}
 \end{equation}
  We may simplify $\Ewv{ED}$ using either (\ref{EDV}) or (\ref{EDV});
equating the results shows that $\alpha\Ewv{E}=\beta\Ewv{D}$, so that we may
write
\begin{eqnarray}
  \Ewv{D}&=&\frac{p}{\beta}\gamma\Ewvi\label{WDV}\,,\\
  \Ewv{E}&=&\frac{p}{\alpha}\gamma\Ewvi\label{WEV}\,,
\end{eqnarray}
 for some constant $\gamma$.  Similarly, $\Ewv{DED}$ can be
simplified using either (\ref{WDED}) or (\ref{EDV}), and this leads to
 \begin{equation}
  \Ewv{DE} = (1-\beta)\Ewv{D}+(1-\alpha)\Ewv{E}+p\gamma\Ewvi\,.\label{WDEV}
 \end{equation}
 Relations (\ref{bulk1})--(\ref{WDEV}) allow the straightforward
computation of all $f_N(\systau)$.

Equation (\ref{bulk1}), when inserted in (\ref{weight}), leads to the
recursion relation (\ref{example_recur}).  From the set of all the
algebraic rules one may similarly construct a whole set of recursion
relations which uniquely specifies the steady state weights.  It is more
convenient, however, to work directly with the operator product.

The algebra is not well defined when $\alpha$ or $\beta$ vanishes, and the
model is not interesting when $p=0$.  However, one may consider 
the random sequential limit
discussed in the introduction, $\alpha = p \widetilde{\alpha}$,
$\beta = p \widetilde{\beta}$, $p \to 0$. Assuming that the operators
$E$, $D$  and vectors $\langle W|$, $|V\rangle$ of our algebra have  
limits $\widetilde E$, $\widetilde D$, $\langle {\widetilde W}|$ and
$|{\widetilde V}\rangle$ under this scaling, and also assuming that 
$\gamma$ has the limit ${\widetilde \gamma}=1$ (as it is true for the
representations we construct in section \ref{representations}) one
finds that (\ref{bulk1}--\ref{WED}) and (\ref{WDV}--\ref{WDEV})
are, in the limit, consequences of the quadratic
algebra of \cite{DEHP}, which reads
\begin{eqnarray}
{\widetilde D}{\widetilde E} &=& {\widetilde D}\, +\, {\widetilde E}\, , \label{dehp1}\\
{\widetilde D}|{\widetilde V}\rangle &=& \frac{1}{\widetilde \beta}
|{\widetilde V}\rangle\, ,
\label{dehp2}\\ \label{dehp3}
\langle {\widetilde W} | {\widetilde E}  &=& \frac{1}{\widetilde
\alpha}\langle {\widetilde W}| \, .
\end{eqnarray}

{\noindent\bf Remark 2.1:} The relations (\ref{bulk1})--(\ref{bulk4}) can be
used to obtain the steady state for our model with periodic boundary
conditions (i.e., on a ring) if (\ref{weight}) is replaced by
$f_N(\systau)={\rm Tr}\Bigl((\prod_{i=1}^{N}(1-\tau_i)E+\tau_iD)\Bigr)\;.$
However, the algebra is not needed in this simple case, because the steady
state is already known \cite{traffic}.
\section{Proof of stationarity}
\setcounter{equation}{0}

In this section we show that the operator algebra (\ref{bulk1})--(\ref{WDED})
may be used to compute the stationary state of the ASEP with parallel
dynamics.  An elementary recursive argument shows that the weights
$f_N(\systau)$ defined by (\ref{weight}) and the constant $\gamma$ introduced
in (\ref{WDV}) and (\ref{WEV}) satisfy $f_N(\systau)/\gamma>0$ for $N\ge1$,
so that (\ref{PN}) defines a probability distribution on the set of all
system configurations.  We must show that this distribution is invariant
under the dynamics.  To avoid consideration of many special cases it is
convenient to first rewrite the algebraic relations satisfied by $D$, $E$,
$\V$, and $\W$ in more unified form. 

Note first that relations (\ref{bulk1})--(\ref{bulk4}),
(\ref{DDEV})--(\ref{WDED}), and (\ref{WDEV}) all involve four factors (from
among $D$, $E$, $\W$, and $\V$) on their left hand sides.  These relations
may be expressed by the single equation
 \begin{equation}
 XDEY =  a(XDY)\,XDY + a(XEY)\,XEY + p\,a(XY)\,XY\, .
\label{blk}
 \end{equation}
 Here $X$ denotes either $\W$, $D$, or $E$ and $Y$ denotes $D$, $E$, or $\V$. 
The coefficient $a(S)$ of a term $S$ ($S=XDY$, $XEY$, or $XY$) on the right
hand side of (\ref{blk}) is determined only by $S$ itself, not the relation
under consideration: 

 \begin{equation}
 a(S)=\left\{\begin{array}{ll}1-p,& \mbox{if $S$ contains $DE$,}\\
   1-\alpha,& \mbox{if $S$ contains $\W E$,}\\
   1-\beta,& \mbox{if $S$ contains $D\V$,}\\
   \gamma,& \mbox{if $S$ is  $\Ewvi$,}\\
   1,& \mbox{otherwise.}\end{array}\right.\label{aS}
 \end{equation}
(Equation (\ref{aS}) represents a slight abuse of notation, since $a(S)$
really depends on the {\it form} of $S$ rather than on the value of $S$,
which may be an operator, a vector, or a scalar; no confusion should
arise.) We obtain another form of the relation (\ref{blk}) by first writing
$XDEY=(1-p)XDEY+pXDEY$ and and then using (\ref{blk}) in the second term:
 \begin{equation}
 XDEY =  a(XDEY)\,XDEY + p\,a(XDY)\,XDY + p\,a(XEY)\,XEY + p^2\,a(XY)\,XY\, .
\label{blki}
 \end{equation}

The relations (\ref{DDV})--(\ref{WED}) and (\ref{WDV})--(\ref{WEV}) can be
similarly unified:
 \begin{eqnarray}
 XD\V &=&  \frac{p}{\beta} a(X\V)\,X\V\;, \label{bdry1} \\
 \W EY &=&  \frac{p}{\alpha} a(\W Y)\,\W Y.
\label{bdry2}
 \end{eqnarray}
 Here, as in (\ref{blk}), $X$ is $\W$, $D$, or $E$ and $Y$ is $D$, $E$, or
$\V$.  A second form of the relations (\ref{bdry1}) and (\ref{bdry2}) is
obtained as was (\ref{blki}), starting from $XFY=(1-\beta)XFY+\beta XFY$
for (\ref{bdry1}) and $XFY=(1-\alpha)XFY+\alpha XFY$ for (\ref{bdry2}):
 \begin{eqnarray}
 XD\V &=&  a(XD\V)\,XD\V + p\,a(X\V)\,X\V \;,\label{bdryi1} \\
 \W EY &=&  a(\W EY)\,\W EY + p\,a(\W Y)\,\W Y\;.
\label{bdryi2}
 \end{eqnarray}

\noindent
 {\bf Remark 3.1:} (i)~The set of terms on the right side of (\ref{blk}) is
obtained by omitting either or both of the middle two factors in $XDEY$,
and the set on the right side of (\ref{blki}) by omitting neither, either,
or both.  Similarly, terms on the right side of (\ref{bdry1}) and
(\ref{bdry2}) arise through the omission of one operator, and those in
(\ref{bdryi1}) and (\ref{bdryi2}) through the omission of zero or one.
(ii)~In (\ref{blki}), (\ref{bdryi1}), and (\ref{bdryi2}) the power of $p$
in each term is the number of operators omitted in obtaining that term; in
(\ref{blk}), it is one less than that number.

We now write down the stationarity condition which
the weights $f_N(\systau)$ must satisfy.  If we imagine for the moment that
our lattice contains two extra boundary sites, $0$ on the left and $N+1$ on
the right, then there are $N+1$ bonds $(i,i+1)$ across which an exchange
might occur during one step in the evolution; here by an ``exchange''
across $(0,1)$ or $(N,N+1)$ we mean the entry or exit, respectively, of a
particle.  Given a fixed configuration $\systau$, let us write
$\A(\systau)$ for the subset of these bonds across which an exchange can
occur in $\systau$ and $\B(\systau)$ for the complementary subset of bonds
across which an exchange might have occurred in arriving at $\systau$ from
some immediate predecessor; the bonds in $\A(\systau)$ correspond to
$\W E$, $DE$, or $D\V$ in the formula (\ref{weight}) for $f_N(\systau)$,
while those in $\B(\systau)$ correspond to $\W D$, $ED$, or $E\V$.  For
$\C\subset\B(\systau)$ write $\systau^{\C}$ for the configuration obtained
from $\systau$ by making the exchanges corresponding to the bonds in $\C$;
the configurations $\systau^\C$, $\C\subset\B(\systau)$, comprise all possible
immediate predecessors of $\systau$.  Then the stationarity condition has
the form
 \begin{equation}
 f_N(\systau) 
    = \sum_{\C\subset\B(\systau)} \pi(\C) f_N(\systau^\C)\,.\label{stat}
 \end{equation}
 Here for any subset $\C$ of $\B(\systau)$, $\pi(\C)$ is the probability
that, in the configuration $\systau^\C$, precisely the set of exchanges
$\C$ (out of the set $\A(\systau^\C)$ of possible exchanges) were in fact
made; thus $\pi(\C)$ is a product of the following factors:
\\
 \hbox to 20pt{\hss$\bullet$\hss}$\alpha$,
   if $(0,1)\in\C$; $(1-\alpha)$, if $(0,1)\in\A(\systau^\C)\setminus\C$;
\\
\hbox to 20pt{\hss$\bullet$\hss}$p$, for each $(i,i+1)\in\C$; 
  $(1-p)$, for each    $(i,i+1)\in\A(\systau^\C)\setminus\C$ ($1\le i\le N-1$);\\
\hbox to 20pt{\hss$\bullet$\hss}$\beta$, if $(N,N+1)\in\C$; 
   $(1-\beta)$, if $(N,N+1)\in\A(\systau^\C)\setminus\C$;
\\
with $\A(\systau^\C)\setminus\C$ denoting the set of bonds belonging to
$\A(\systau^\C)$ but not to $\C$.

To verify (\ref{stat}) we will use the relations above to transform each
side to a common form.  We need one more concept.  The configuration
$\systau$ may be divided into blocks of successive zeros and ones; let
$\E(\systau)$ denote the set of such blocks and for $\F\subset \E(\systau)$
let $\systau_\F$ denote the configuration obtained from $\systau$ by
omitting one operator from each block in $\F$.

We first consider the left hand side of (\ref{stat}).  In the expression
(\ref{weight}) for $f_N(\systau)$ we apply (\ref{blki}) to each factor $DE$,
(\ref{bdryi1}) to each $D\V$, and  (\ref{bdryi2}) to each $\W E$, if these
occur.  The resulting sum over all ways of omitting zero, one, or two
operators at each of these bonds (see Remark~3.1.i) is equivalent to a sum
over all ways of omitting zero or one operators from each block in
$\systau$, so that from (\ref{aS}) and Remark~3.1.ii we obtain
 \begin{equation}
 \sum_{\F\subset\E(\systau)} p^{|\F|} (1-\alpha)^{x(\F)}(1-p)^{y(\F)}
  (1-\beta)^{z(\F)} f_{N-|\F|}(\systau_\F), \label{lhs}
 \end{equation}
 where $|\F|$ is the number of elements in $\F$ and $x(\F)$, $y(\F)$, and
$z(\F)$ count respectively the number of factors $\W E$, $DE$, and $D\V$ in
the expression (\ref{weight}) for $f(\systau_\F)$.  In the special case
$f_{N-|\F|}(\systau_\F)=\Ewvi$, which can occur only if
$\systau=1\,0\,1\,0\,1\,0\,\dots$ or $\systau=0\,1\,0\,1\,0\,1\,\dots$,
there will be an additional factor of $\gamma$. It is important
here that the coefficients $a(S)$ in (\ref{blki}), (\ref{bdryi1}), and
(\ref{bdryi2}) depend only on $S$, so that the coefficients in (\ref{lhs})
depend only on $\F$ and in particular are independent of the order in which
the relations (\ref{blki}), (\ref{bdryi1}), and (\ref{bdryi2}) are applied
at different bonds.  Similar comments apply to the expansions using
(\ref{blk}), (\ref{bdry1}), and (\ref{bdryi2}) which we will perform below.

Consider now a weight $f_N(\systau^\C)$ which occurs in the sum on the
right hand side of (\ref{stat}).  Each bond in $\C$ corresponds in the
expression (\ref{weight}) for this weight to a factor $DE$, $D\V$, or
$\W E$; we apply (\ref{blk}), (\ref{bdry1}), or (\ref{bdry2}) to these
factors.  The result will be of the form
 \begin{equation}
 f_N(\systau^\C)= \sum_\F p^{|\F|-|\C|}\lambda(\F)f_{N-|\F|}(\systau_\F).
    \label{fnc}
 \end{equation}
Since each of the relations we are using involves the omission of one or
two operators (see Remark~3.1.i), the $\F$ occurring in (\ref{fnc}) will be
those which, for each bond in $\C$, contain one or both of the blocks
abutting on this bond.  The total number of operators omitted is $|\F|$
and the number of times one of the relations was used is $|\C|$, so that
the factor $p^{|\F|-|\C|}$ is obtained directly from Remark~3.1.ii.  The
coefficient $\lambda(\F)$ is a product which contains a factor $p/\beta$ if
(\ref{bdry1}) was used (i.e., if $(0,1)\in\C$) and $p/\alpha$ if
(\ref{bdry2}) was used ($(N,N+1))\in\C$).  It also contains factors arising
from (\ref{aS}): a factor of $1-p$, $1-\beta$, or $1-\alpha$ for each
application of (\ref{blk}), (\ref{bdry1}), or (\ref{bdry2}) which preserves
or generates a factor $DE$, $D\V$, or $\W E$ respectively, and a factor
$\gamma$ if $f_{N-|\F|}(\systau_\F)=\Ewvi$.

When (\ref{fnc}) is inserted into the right side of (\ref{stat}) the double
sum over $\C$ and $\F$ becomes a sum over {\it all} subsets
$\F\subset\E(\systau)$,
 \begin{equation}
 \sum_{\F\subset\E(\systau)} \pi(\C)p^{|\F|-|\C|}\lambda(\F)
   f_{N-|\F|}(\systau_\F),
    \label{rhs}
 \end{equation}
 since, given any $\F$, one may identify the corresponding $\C$ in
(\ref{fnc}) as the set of bonds in $\B(\systau)$ for which no block
belonging to $\F$ abuts on $\C$.  It is straightforward to complete the
proof by verifying that $p^{|\F|-|\C|}\pi(\C)\lambda(\F)$ is precisely the
coefficient of $f_{N-|\F|}(\systau_\F)$ in (\ref{lhs}), and hence that
(\ref{lhs}) and (\ref{rhs}) agree.  In particular, $\pi(\C)\lambda(\F)$
contains a factor $p^{|\C|}$, that is, one factor of $p$ for each bond in
$\C$; for internal bonds these factors are present in $\pi(\C)$, while if
$(0,1)\in\C$ then $\pi(\C)$ contains a factor $\alpha$ and $\lambda(\F)$ a
factor $p/\alpha$ (the argument for $(N,N+1)\in\C$ is similar).  If
(\ref{lhs}) contains a factor $(1-\alpha)$, that is, if $f(\systau_\F)$
contains $\W E$, then $\pi(\C)$ contains this factor if $(0,1)\notin\C$ and
$\lambda(\F)$ if $(0,1)\in\C$; the factors of $(1-p)$ and $(1-\beta)$ in
(\ref{lhs}) are accounted for similarly.
\section{Reduction to a Quadratic Algebra}
\label{reduction}
\setcounter{equation}{0}

In this section we show that the quartic algebraic rules 
(\ref{bulk1})--(\ref{WED})
can be reduced to quadratic rules by making a convenient
choice for the operators involved. The trick is to write
\begin{eqnarray}      
{ D} = \left( \begin{array}{cc}      
 D_1&0\\      
 D_2&0
 \end{array}      
 \right) \;,\;\;\;
{ E} = \left( \begin{array}{cc}
 E_1&E_2\\
 0&0
 \end{array}      
 \right) \;, \label{DE}
 \end{eqnarray}
 where $D_1$, $D_2$, $E_1$, and $E_2$ are 
matrices of arbitrary (in general infinite)
dimension; that is, $D$ and $E$ are 
written as rank four tensors with two indices
of (possibly) infinite dimension and the other two indices of dimension two. 
Correspondingly, we write $\langle { W}|$ and $| { V} \rangle$ in the form
 \begin{eqnarray}
\langle { W}| = (\ \langle W_1|,\ \langle W_2|\ )
  \;,\;\;\; 
| { W}\rangle =
 \left( \begin{array}{c}| V_1 \rangle \\ | V_2 \rangle \end{array} \right) \;,
  \label{WV}
\end {eqnarray}
 where $\langle W_1|$, $\langle W_2|$, $| V_1 \rangle$, 
and $| V_2 \rangle $ are
vectors of the same dimension as $D_1$ and $E_1$.  We will show that the
operators and vectors so defined satisfy the algebra of
section~\ref{fundamental} if $D_1$, $E_1$, $\langle W_1|$, and
$|V_1\rangle$ satisfy the quadratic relations
\begin{equation}
D_1 E_1 = (1-p)\left[ D_1 +E_1 +p  \right] \;, \label{D1E1con}
\end{equation}\begin{equation}
D_1 | V_1 \rangle = \frac{p(1-\beta)}{\beta} |V_1\rangle \;\;\;,\;\;\;
\langle W_1|E_1 =  \langle W_1| \frac{p(1-\alpha)}{\alpha}, \label{D1V1}
\end{equation}
 and $D_2$, $E_2$, $\langle W_2|$, and $|V_2\rangle$ satisfy 
\begin{equation}
E_2 D_2 = p \left[ D_1 +E_1 +p  \right], \label{E2D2con}
\end{equation}\begin{equation}
E_2 | V_2 \rangle = p|V_1\rangle \;\;\;,\;\;\;
\langle W_2|D_2 = \langle W_1|p. \label{E2V2}
\end{equation}

We now verify that (\ref{DE})--(\ref{E2V2}) imply (\ref{bulk1})--(\ref{WDEV}). 
First, by substituting (\ref{DE}) into the bulk relations
(\ref{bulk1})--(\ref{bulk4}) one finds that to satisfy the latter equations it
is sufficient that
\begin{eqnarray}
E_1D_1E_1+E_2D_2E_1 &=& (1-p)E_1D_1 + (1-p)E_2D_2
+E_1E_1 + pE_1\;,
\label{bulk5} \\
E_1D_1E_1D_1+E_2D_2E_1D_1+
E_1D_1E_2D_2+E_2D_2E_2D_2\hspace{-2.33in} 
&&\\
 &=& E_1D_1D_1 +E_2D_2D_1 +
E_1E_1D_1 + E_1E_2D_2 +pE_1D_1 +pE_2D_2\;, \nonumber \\
 D_1E_1 &=&(1-p)E_1 +(1-p)D_1 +p(1-p)\;,\\
D_1E_1D_1+D_1E_2D_2 &=& (1-p)E_1D_1 +(1-p) E_2D_2 +D_1D_1 +pD_1 \;.
\label{bulk8}
\end{eqnarray}
These relations follow from (\ref{D1E1con}) and (\ref{E2D2con}).  For
example, the left hand side of (\ref{bulk5}) becomes
 \begin{equation}
(1-p)E_1 \left[D_1 +E_1 +p \right]
+ p \left[D_1 +E_1 +p  \right]E_1
= (1-p)E_1D_1 +pD_1E_1 +E_1E_1 +pE_1\;,
 \end{equation}
 which another use of (\ref{D1E1con}) and (\ref{E2D2con}) 
shows to be equal to the
right hand side.  Similarly, with (\ref{DE}) and (\ref{WV}), the relations
(\ref{DDV})--(\ref{WED}) involving two sites next to the right hand boundary
follow from
\begin{eqnarray}
\left( \begin{array}{c}  D_1 D_1 |V_1\rangle\\  D_2 D_1 |V_1\rangle 
\end{array} \right)
&=&\frac{p(1-\beta)}{\beta}\left( \begin{array}{c}  D_1 |V_1\rangle  \\
D_2 |V_1\rangle \end{array} \right)\\
E_1 D_1 |V_1\rangle  +E_2 D_2 |V_1\rangle &=& \frac{p}{\beta} E_1|V_1\rangle
         +\frac{p}{\beta} E_2|V_2\rangle\;,
\end{eqnarray}
 and these equations are in turn implied by (\ref{D1V1}) and (\ref{E2V2}). 
The conditions (\ref{DDEV}) and (\ref{EDEV}), involving three sites next to the
right boundary, are obtained similarly, although the manipulations involved
become rather tedious.  Relations at the left boundary are obtained from
symmetry considerations, completing the verification.

In the remainder of the paper, we will consider only representations of the
quadratic algebra having the form of (\ref{DE}) and (\ref{WV}).  As we now
discuss, computations in this representation are simplified by the fact
that all quantities of physical interest may be expressed in terms of
$D_1$, $E_1$, $\langle W_1|$, and $|V_1\rangle$.  This also leads to
connections with other updating procedures for the ASEP (see
section~\ref{mappings}).

Let us first derive such an expression for the normalization constant $Z_N$,
given by
 \begin{equation}
Z_N = \langle { W}|{ C}^N | { V} \rangle \;,
 \end{equation}
where
 \begin{equation}
{ C} = { D} + { E}=
\left( \begin{array}{cc}
C_1& E_2\\
D_2& 0
\end{array}
\right)\;
 \end{equation}
and $C_1=D_1+E_1$. Now, 
 \begin{equation}
{ C}^N =
\left( \begin{array}{cc}
G(N)& G(N-1)E_2\\
D_2G(N-1)& D_2G(N-2)E_2
\label{CN}
\end{array}
\right)\;,
 \end{equation}
where by convention $G(-1)=0$ and
 \begin{equation}
G(N) = \sum_{n=0}^{N} K^{N-n} (-p)^n 
\label{Gdef}
 \end{equation}
 for $N\ge1$, with
 \begin{equation}
K = (C_1 + p)\;.
\label{Kdef}
 \end{equation}
This may be proven by first checking the case $N=0$
($G(0)=\thickone$) and then verifying the recursion
$G(N+1) =C_1G(N)+E_2D_2G(N-1)
= (K-p) G(N) +p K G(N-1) $.  Note that 
 \begin{equation}
  G(N)+pG(N-1)=K^N.
 \label{GandK}
 \end{equation}
From (\ref{CN}), (\ref{GandK}), and the action (\ref{E2V2}) of $E_2$ and $D_2$
on the boundary vectors, we have
 \begin{eqnarray}
\langle { W}|{ C}^n 
&=&(\ \langle W_1| K^n,\ \langle W_1|K^{n-1}E_2\ ) \label{WCn} \;, \\
{ C}^n | { V} \rangle 
&=&
 \left( \begin{array}{c}K^n| V_1 \rangle \\ D_2
 K^{n-1}| V_1 \rangle \end{array} \right) \;, \label{CnV}
 \end{eqnarray}
which leads to
 \begin{equation}
Z_N =z_N +pz_{N-1} \;,
\label{Z2}
 \end{equation}
where
 \begin{equation}
z_n=\langle W_1|K^n |V_1 \rangle . \label{littlez}
 \end{equation}

An important quantity in determining the phase diagram
is the current $J_N$,  which is the probability that 
a particle passes through a particular bond in a particular time step.
It is given by any one of three equivalent expressions:
 \begin{equation}
J_N = \alpha \langle (1-\tau_1) \rangle
=p \langle \tau_i(1-\tau_{i+1}) \rangle = \beta \langle \tau_N \rangle,
 \end{equation}
 where $1<i<N$; the second of these may be written using the
algebra as 
 \begin{equation}
J_N =  p\frac{\langle { W}|{ C}^{i-1}{ DE}{ C}^{N-i-1} 
   | { V} \rangle}{Z_N } \;. \label{currfrac}
 \end{equation}
 Now (\ref{WCn}) and (\ref{CnV}) yield
 \begin{eqnarray}
\langle { W}|{ C}^nD  &=&
(\ \langle W_1| K^{n-1}[D_1 + p],\ 0\ ) \;,
 \label{WCnD} \\
E { C}^n | { V} \rangle 
&=&
 \left( \begin{array}{c}[E_1 + p ]K^{n-1}| V_1 \rangle \\ 
0 \end{array} \right)\;, \label{ECnV}
 \end{eqnarray}
 and the algebraic rule (\ref{D1E1con}) implies that
 \begin{equation}
[D_1 + p][E_1 + p ] = K,
 \end{equation}
 so that from (\ref{currfrac}), (\ref{Z2}), and (\ref{littlez}),
 \begin{equation}
J_N = \frac{pz_{N-1}}{z_N +pz_{N-1}}.
\label{Jeq}
 \end{equation}
 This expression again involves only the matrices $E_1$ and $D_1$ and
vectors $\langle W_1 |$ and $| V_1 \rangle$.

 We may similarly express the one-point correlation function or density
profile,
 \begin{equation}
\langle \tau_i \rangle_N
= \frac{1}{Z_N} \langle { W} | { C}^{i-1}
{ D} { C}^{N-i} | { V} \rangle ,
 \end{equation}
 in terms of  $E_1$, $D_1$, $\langle W_1 |$, and $| V_1 \rangle$:
 \begin{equation}
\langle \tau_i \rangle_N=\frac{
 \langle W_1 | K^{i-1} (D_1+p) K^{N-i} |  V_1 \rangle}
{ z_N +p z_{N-1}}
 \label{density} 
 \end{equation}
 where we have employed (\ref{WCnD}) and (\ref{CnV}).  
Similar expressions are possible
for higher correlation functions; for example, the two-point correlation
function $\langle \tau_i (1- \tau_j) \rangle_N$ is given, from
(\ref{CN}), (\ref{WCnD}) and (\ref{ECnV})
  by
 \begin{eqnarray}
\langle \tau_i (1- \tau_j) \rangle_N
&=&
\frac{1}{Z_N} \langle { W} | { C}^{i-1}
{ D} { C}^{j-i-1}{ E} { C}^{N-j} | { V} \rangle \nonumber \\
&=&
\frac{1}{Z_N} \langle W_1 | K^{i-1} (D_1 +p) G(j-i-1) (E_1+p) K^{N-j}
|  V_1 
\rangle\, .
\label{twopoint}
 \end{eqnarray}
 Because $G(n)$ is an alternating sum, (\ref{twopoint}) is more complicated
than the corresponding expression in the random sequential case; this
reflects the fact that stronger correlations exist under parallel dynamics
than under random sequential dynamics. 

In section~\ref{combinatorial} we will give a more explicit formulae
for $\langle\tau_i\rangle_N$; see (\ref{tauexact}).
\section{Relations with other models}
\label{mappings}
\setcounter{equation}{0}

The reduction to a quadratic algebra and the expressions for the current and
correlation functions, derived in section~\ref{reduction}, lead to relations
between the ASEP with parallel updating and the same model with certain other
discrete-time updating procedures.  The procedures in question can in fact be
defined for more general site variables and for any local dynamical rules
which assign to each configuration of a pair of sites at time $t$
a new configuration at time $t+1$ with some given probability, 
and similarly to each configuration on the
leftmost or rightmost site a new configuration.  
We recall these procedures briefly; see \cite{RSSS} for precise
definitions.  In the {\it ordered sequential\/} update sites are updated one
at a time, starting the right end of the system and proceeding sequentially
to the left end (backward ordered), or vice versa
(forward ordered).  In the {\it sublattice parallel\/} update,
all site pairs $i,i+1$ with $i$ even are updated at one time step and and all
such pairs with $i$ odd at the next time step.  Let us denote these
procedures by the symbols $T_{\leftarrow}$, $T_{\rightarrow}$ and $T_{\sp}$
(more precisely $T_{\leftarrow}$, $T_{\rightarrow}$ and $T_{\sp}$ are
the transfer matrices for the different procedures, however since in this
discussion we do not require any  properties of the transfer matrices,
we do not give  detailed definitions).

It can be shown \cite{rajsch} using the matrix product formalism that in
general the procedures $T_{\leftarrow}$, $T_{\rightarrow}$ and $T_{\sp}$
lead to stationary states which
may be regarded as physically equivalent.
In particular, the current is independent of the update procedure; we
write $J^{\#}$ for this common value:
 \begin{equation}
J_N^{\#}=J_N^{\leftarrow}=J_N^{\rightarrow}=
J_N^{\rm \sp}\;. \label{currents}
 \end{equation}
The density profiles are also closely related (we use the notation of the
ASEP, but a corresponding result holds in general \cite{rajsch}):
 \begin{equation}
 \langle \tau_i\rangle_N^{\sp}
    =\left\{\begin{array}{ll}
    \langle \tau_i\rangle_N^{\leftarrow},& i \hbox{ even,}\\
    \langle \tau_i\rangle_N^{\rightarrow},& i \hbox{ odd,}
 \end{array}\right. 
\qquad\qquad
 \langle \tau_i\rangle_N^{\sp*}
    =\left\{\begin{array}{ll}
    \langle \tau_i\rangle_N^{\leftarrow},& i \hbox{ odd,}\\
    \langle \tau_i\rangle_N^{\rightarrow},& i \hbox{ even,}
 \end{array}\right. \label{densities}
 \end{equation}
where $\langle\tau_i\rangle_N^{\sp*}$ is the density for $T_{\sp}$ after
the first (even) sublattice has been updated.  Similar relations hold for some
higher order correlation functions.

More is known in the special case of the ASEP, to which we limit our
discussion in what follows, based on simpler realizations of the matrix
product ansatz in that case \cite{hinri,work,honi,RSSS}. 
In particular there is another relation between density profiles, 
 \begin{equation} \label{conn}
  \langle \tau_i\rangle_N^{\rightarrow}
   =\langle \tau_i\rangle_N^{\leftarrow}-J_N^{\#},
 \end{equation}
 which, with (\ref{densities}), shows that any one of
$\langle\tau_i\rangle_N^{\leftarrow}$,
$\langle\tau_i\rangle_N^{\rightarrow}$, and $\langle\tau_i\rangle_N^{\sp}$
determines the others. Moreover, the asymptotic current
$\lim_{N\to\infty}J_N^{\#}$ (and therefore the phase diagram) is known
\cite{honi}, and so are the density profiles and correlation functions
in the case $p=1$ \cite{hinri,work}.

A fully parallel updating procedure is not naturally defined for arbitrary
local dynamical rules, due to the possibility of conflict when the rules are
applied simultaneously to pairs overlapping of sites, but for the ASEP such a
fully parallel procedure, which we will denote by $T_{\parallel}$, has been
defined in section~\ref{fundamental}.  We now want to relate this model to
the ASEP with update procedures $T_{\rightarrow}$, $T_{\leftarrow}$, and
$T_{\sp}$, considered above; for the ASEP with parallel update we
work with the reduced version of the algebra
established in section~\ref{reduction}, in which everything is expressed in
terms of the matrices $E_1$ and $D_1$ and vectors $\langle W_1 |$ and $| V_1
\rangle$, which satisfy the algebraic relations (\ref{D1E1con}, \ref{D1V1}). 
The idea is to introduce new operators $e$ and $d$ by 
\begin{equation}
d=D_1 \;\;\;,\;\;\; e=E_1+p \;,
\end{equation}
so that $K=E_1+D_1+p=e+d$.  From (\ref{D1E1con}, \ref{D1V1}), the new
matrices satisfy
 \begin{equation}\label{bulk2bis}
de=d+(1-p){e}
 \end{equation}
and
 \begin{equation}\label{boundaries2}
d|V_1\rangle =\frac{p(1-\beta)}{\beta}|V_1\rangle  \quad,\quad
\langle W_1|{e}=\langle W_1|\frac{p}{\alpha} \quad.
 \end{equation}
Surprisingly, these equations are precisely the algebraic relations
\cite{work} for the matrix product solution of the ASEP with forward
updating, that is, the steady state weight for the configuration $\tau$
with updating $T_{\rightarrow}$ is given by
 \begin{equation}
 \langle W_1|(\prod_{i=1}^{N}(1-\tau_i)e+\tau_id)|V_1\rangle\;.
 \end{equation}
Writing $c=e+d$ we see that the normalizing factor for $T_{\rightarrow}$ is
 \begin{equation}\label{norm}
  \langle W_1|c^N|V_1\rangle = \langle W_1|K^N|V_1\rangle = z_N \;,
 \end{equation}
 a constant already introduced in (\ref{littlez}), and the current is
 \begin{equation}\label{currsequ}
  J_N^{\#}=\alpha\frac{\langle W_1|e{c}^{N-1}|V_1\rangle}{z_N}=p\frac{z_{N-1}}{z_N}\;,
 \end{equation}
The density profile is given by
 \begin{equation}\label{denssequ}
 \langle\tau_i\rangle^{\rightarrow}
  =\frac{\langle W_1|c^{i-1}dc^{N-i}|V_1\rangle}{z_N}
  =\frac{\langle W_1|K^{i-1}D_1K^{N-i}|V_1\rangle}
   {\langle W_1|K^N|V_1\rangle}\; . \label{dprightarrow}
 \end{equation}

These formulae imply certain simple relations between physical quantities in
the parallel and the ordered sequential ASEP.  From (\ref{currsequ}) and the
formula (\ref{Jeq}) for the current in the ASEP with update $T_{\parallel}$,
$J_N=pz_{N-1}/(z_N+pz_{N-1})$, we have
 \begin{equation}\label{currtrafo}
 J_N=\frac{J_N^{\#}}{1+J_N^{\#}}\;.
 \end{equation}
 In particular, $J_N<J_N^{\#}$, a result which is intuitively
clear.  Similarly, comparing (\ref{denssequ}) with the formula
(\ref{density}) for the density profile for the parallel ASEP leads to
 \begin{equation}\label{denstrafo}
\langle\tau_i\rangle_N=
\frac{\langle\tau_i\rangle^{\rightarrow}_N
+J_N^{\#}}{1+J_N^{\#}}=
\frac{\langle\tau_i\rangle^{\leftarrow}_N}
{1+J_N^{\#}}\, ,
 \end{equation}
 where we have also used (\ref{conn}).  It is also possible to derive
similar formulae for the higher correlation functions
although these are not so simple. For example, the
two-point function (\ref{twopoint}) for $T_{\parallel}$ can be written as
an alternating sum over two-point functions of $T_{\rightarrow}$:
 \begin{equation}\label{twopttrafo}
\langle\tau_i(1-\tau_j)\rangle_N=
\frac{1}{Z_N} \sum_{n=0}^{j-i-1}
(-p)^n
\left[
\langle\tau_i(1-\tau_{j-n})\rangle_{N-n}^{\rightarrow} z_{N-n}
-p\langle\tau_i\rangle_{N-n-1}^{\rightarrow} z_{N-n-1}
\right]\;,
 \end{equation}
 although this formula
appears to be too complicated to be of practical use
in obtaining $ \langle\tau_i(1-\tau_j)\rangle_N$  from
$\langle\tau_i(1-\tau_j)\rangle_N^{\rightarrow}$ or vice versa.
 We emphasize that the formulae (\ref{currtrafo}), (\ref{denstrafo}), and
(\ref{twopttrafo}) hold for all $\alpha$, $\beta$, and $p$.  Of course, using
the formulae established earlier in this section, we obtain relations
for the one- and two-point functions with updates
 $T_{\leftarrow}$ and $T_{\sp}$. 

The relation (\ref{currtrafo}) for the currents and the relation for the
bulk densities which follows from (\ref{denstrafo}) were already
conjectured in \cite{RSSS} (see Table~1 in that reference).

These relations between different models have various consequences.  On
the one
hand, established results for $T_{\leftarrow}$, $T_{\rightarrow}$ and
$T_{\sp}$ serve as a check for the results we will derive later; this applies
to the asymptotic values of the currents and hence to the phase diagram which
we derive in section~\ref{phasediagram}, and to some of the results in the
special case $p=1$.  

On the other hand, and more importantly, most of the results that we will
derive in later sections---explicit representations of the algebra, some of
the detailed results for the case $p=1$, asymptotics of density profiles, and
finite volume formulas for current and density profiles---apply to all of the
update procedures $T_{\parallel}$, $T_{\leftarrow}$, $T_{\rightarrow}$, and
$T_{\sp}$.  We will state these in terms of $T_{\parallel}$, since that
procedure is the main focus of this paper, but results for other updates are
easily obtained via the formulae of this section.

 Finally, we want to point out a further connection to another known model,
the ASEP with random sequential update on a ring with a second class
particle \cite{DJLS,mallick}. In the ASEP with two species of particles a
first class particle hops onto a vacant site to its right with rate one and
exchanges positions with a second class particle to its right with rate
$r$, and a second class particle hops onto a vacant site to its right with
rate $s$.  Let us summarize some known results for this model.  The
stationary state on a ring can be written as a matrix product state
\cite{DJLS}; the corresponding matrix algebra is given by
(\ref{dehp1})--(\ref{dehp3}), where now the matrix for
first class particles is ${\widetilde D}$,  the  matrix for holes is
${\widetilde E}$, the matrix
for the second class particle is
$|{\widetilde V}\rangle\langle {\widetilde W}|$, and
${\widetilde \alpha}=r$, ${\widetilde \beta}=s$.  Since we are considering
a closed system, it is convenient to work in a grand canonical ensemble
\cite{DJLS} with fugacity $x$ for first class particles. When the system
contains just one second class particle, the density of first class
particles $i$ sites ahead of the second class particle in a ring of $N$
sites is given by
 \begin{equation}
  \langle\tau_i\rangle_N^\scp
   =x \frac{\langle {\widetilde W}| {\widetilde C}^{i-1}
{\widetilde D} {\widetilde C}^{N-i-1} |{\widetilde V}\rangle }
 {\langle{\widetilde W}| {\widetilde C}^{N-1}|{\widetilde V}\rangle} \, ,
 \label{scpdensity}
\end{equation}
 where ${\widetilde C}=x{\widetilde D}+{\widetilde E}$, with
$x$ the fugacity.

Now we can  map the algebra (\ref{bulk2bis}), (\ref{boundaries2})
onto that of (\ref{dehp1})--(\ref{dehp3}) by defining
 \begin{equation}
(1-p){\widetilde D}=d\, , \quad {\widetilde E}=e \,,
\label{mappingmallick1}
  \end{equation}
 and
 \begin{equation}
  {\widetilde \alpha}=\frac{p}{\alpha},\quad
  {\widetilde \beta}=\frac{\beta(1-p)}{p(1-\beta)}\,. 
 \label{mappingmallick2}
 \end{equation}
 Then $c=e+d$ is equal to $\widetilde C\big|_{x=1-p}$, so that
one can obtain the density profile for the ASEP with $T_{\rightarrow}$, and
hence for the ASEP with $T_{\parallel}$, from the grand-canonical density
profile (\ref{scpdensity}) seen from the second class particle:
comparing (\ref{dprightarrow}) and (\ref{scpdensity}) yields
 \begin{equation}
  (1-p)\langle\tau_i\rangle^{\rightarrow}_N\Big|_{\alpha,\beta,p}
  = \langle\tau_i\rangle_{N+1}^\scp
    \Big|_{r=\alpha/p,\,s=\beta(1-p)/p(1-\beta),\,x=1-p}.
 \end{equation}
  Other quantities in the models can be related similarly.  When $r=s=1$,
$\langle\tau_i\rangle_N^\scp$ is known exactly for both finite and infinite
systems \cite{DJLS}, yielding $\langle\tau_i\rangle_N^\rightarrow$ for the
case $\alpha=\beta=p$.  The single second-class particle model for general
$r$ and $s$ has been studied \cite{mallick} in the canonical ensemble of a
fixed number of particles on the ring; to the extent to which
the canonical and grand canonical ensembles
are equivalent in the large $N$ limit, these results should correspond with our
results for the model with parallel update.
We checked that indeed the phase
diagram we will derive in section~\ref{phasediagram} translates to the
correct phase diagram for the model of \cite{mallick}.
\section{Representations of the quadratic algebra}
\label{representations}
\setcounter{equation}{0}

In this section we discuss several explicit representations of the quadratic
algebra (\ref{D1E1con}), (\ref{D1V1}).  The situation is like that for other,
similar matrix product algebras: finite dimensional representations exist for
a few special parameter values, and some representation, typically infinite
dimensional, exists for all values. 

In \cite{RSSS} it was observed that for parameter values on the line
 \begin{equation}\label{specialline}
(1-\alpha)(1-\beta)=1-p\, ,
  \end{equation}
 the weight of a configuration in the stationary state could be written as a
product of factors corresponding to clusters of length two.  Thus, we expect
a simplification of our algebraic relations (\ref{D1E1con})--(\ref{E2V2}) along
(\ref{specialline}).  Indeed, for this special case the matrices
$E_1,D_1,E_2,D_2$ and the vectors $\langle W_1 |,\langle W_2|,| V_1 \rangle,|
V_2 \rangle$ can be chosen to be scalars:
\begin{eqnarray}
D_1=p\frac{1-\beta}{\beta}&,&E_1=p\frac{1-\alpha}{\alpha},\\
D_2=&E_2&=p\nu,\\
|V_1\rangle =&\langle W_1|&=\nu,\\
|V_2\rangle =&\langle W_2|&=1, 
\end{eqnarray}
 where $\nu=\sqrt{{p}/{\alpha\beta}}$, so that $E$ and $D$ are $2\times 2$
matrices.

In section~\ref{pequalone} we will treat the case $p=1$ (with $\alpha$ and
$\beta$ arbitrary), in which one can choose a two dimensional representation
of the $E_1$, $D_1$ algebra.  Here, the bulk dynamics is completely
deterministic.  However, the physics is still interesting \cite{TE}, since
$\alpha$ and $\beta$ can induce different phases.  The algebra is
sufficiently simple that we can derive explicit formulae for quantities such
as the fluctuations in the number of particles in the finite system. 

Finite dimensional representations of the
matrix product algebra (\ref{D1E1con}), (\ref{D1V1})
exist only along the line (\ref{specialline}) and when
$p=1$. This can be seen by using the mapping 
(\ref{mappingmallick1}), (\ref{mappingmallick2}) in section
\ref{mappings}, because it has been proven \cite{DEHP} that 
the matrices in (\ref{dehp1})--(\ref{dehp3}) have to be infinite-dimensional
except in the case ${\widetilde
\alpha} +{\widetilde \beta} =1$, which is by (\ref{mappingmallick2})
equivalent to condition (\ref{specialline}). 
Note that the mapping (\ref{mappingmallick1}) and (\ref{mappingmallick2})
is not well-defined for $p=1$; however, in
this case we know that there is a finite dimensional representation,
because we construct it (see section \ref{pequalone}).

Let us now turn to the case of general $p,\alpha,\beta$, in which the
representations must be infinite dimensional.  Such representations are of
use both as a calculational tool and also to demonstrate that non-trivial
representations of the algebra actually do exist.  It can be shown by direct
calculation that the following expressions
satisfy (\ref{D1E1con})--(\ref{E2V2}):
 \begin{eqnarray}  
\langle W_1|= \left( \begin{array}{ccccc}      
 1,&      
 0,&      
 0&      
 .&.      
 \end{array} \right) \; ,      \;\;
|V_1\rangle= \left( \begin{array}{ccccc}      
 1,&
 0,&      
 0&      
 .&.      
 \end{array} \right)^{\mbox{t}} \;,
\label{W1V1}      \\
{D_1} = \left( \begin{array}{llllll}      
 p(1-\beta)/\beta&b&0&0&. &\hspace{0.2in}.\\      
 0&(1-p)&(1-p)^{1/2}&0&& \\      
 0&0&(1-p)&(1-p)^{1/2}&&\\      
 0&0&0&(1-p)&. &\\      
 . &&&&. &\hspace{0.2in}. \\      
 . &&&&&\hspace{0.2in}.      
 \end{array}      
 \hspace{0.2in} \right)     \;,
 \label{D1}\\
{{E_1}} = \left( \begin{array}{llllll}      
 p(1-\alpha)/\alpha&0&0&0&. &\hspace{0.2in}.\\      
 b&(1-p)&0&0&& \\      
 0 &(1-p)^{1/2}&(1-p)&0&&\\      
 0 &0&(1-p)^{1/2}&(1-p)&&\\      
 . &&&. &.& \\      
 .&&&&.&\hspace{0.2in}.      
 \end{array}      
 \hspace{0.2in} \right)       \;,
\label{E1} 
\end{eqnarray}      
where
 \begin{eqnarray}
b^2&=& \frac{p}{\alpha \beta}\left[ (1-p)-(1-\alpha)(1-\beta)\right]
 \label{bsq}
 \end{eqnarray}
 and
 \begin{equation}
E_2 = g D_1\; ,\;\;\;\; D_2= g E_1 \;,
 \end{equation}
 \begin{equation}
|V_2\rangle= \beta/(g (1-\beta))|V_1\rangle \;,\;\;\;\;
\langle W_2|= \langle W_1| \alpha/(g (1-\alpha))\;,
 \end{equation}
 where $g = \sqrt{p/(1-p)}$.  These representations are not defined
when $p=1$, however, as explained above, we will treat this case in
section~\ref{pequalone}.  Note that for $(1-\alpha)(1-\beta)=1-p$ we have
$b^2=0$, so that the $(1,1)$ elements of the matrices decouple from the rest
and we are left with $2\times 2$ operators ${ D,E}$ (see also 4.2).  This
proves a conjecture of \cite{RSSS} for the structure of the 
steady state.  In the limit $p \to 0$ one recovers the 
representations (36) and (37) of \cite{DEHP}. 
\section{The case $p=1$}
\label{pequalone}
\setcounter{equation}{0}

In the case $p=1$ the hopping of  particles
in the
bulk is deterministic;
the only source of randomness comes from the parameters $\alpha$
and $\beta$. We shall see that $\alpha$ and $\beta$ can induce
a high density phase ($\alpha\ge\beta$) and a low density phase ($\beta\ge\alpha$).
For $\alpha=\beta$, these two phases coexist (see below). Note that 
the recursion relation following
from the algebraic rule (\ref{bulk3})
implies that the weight for configurations containing
a particle-particle-hole-hole string is exactly zero. This is also 
immediately evident from the dynamical rules, 
since these substrings can never be created (they are ``Gardens of
Eden'' that can never be entered once left). 

One can 
choose $2\times 2$ representations
for the operators $E_1$ and $D_1$, so that $E$ and $D$ are
$4\times 4$ matrices. In particular, one can verify 
that the following explicit representations satisfy (\ref{D1E1con})--(\ref{E2V2}): 
\begin{eqnarray}      
\hspace*{0.3in}\langle W_1|= \left( \begin{array}{cc}      
 1,&      
 0      
 \end{array} \right) \;\; ,      \;\;
|V_1\rangle= \left( \begin{array}{cc}      
 1,&
 0      
 \end{array} \right)^{\mbox{t}} \;\; , 
\end{eqnarray}\begin{equation}
{D_1} = \left( \begin{array}{cc}      
 (1-\beta)/\beta&-c\\      
 0&0      
 \end{array}      
 \hspace{0.0in} \right) \;\; \quad,\quad      
\nonumber{{E_1}} = \left( \begin{array}{cc}      
 (1-\alpha)/\alpha&0\\      
 c&0      
 \end{array}      
 \hspace{-0.0in} \right) \;\; ,      
\end{equation}      
\begin{equation}      
\langle W_2|= \left( \begin{array}{cc}      
 1/a,&      
 0      
 \end{array} \right) \;\; ,      \;\;
|V_2\rangle= \left( \begin{array}{cc}      
 1/a,&
 0      
 \end{array} \right)^{\mbox{t}} \;\; ,     
\end{equation}\begin{equation}
{E_2} = \left( \begin{array}{cc}      
 a&-c\\      
 0&1\\      
 \end{array}      
 \hspace{0.0in} \right) \;\; \quad,\quad      
{{D_2}} = \left( \begin{array}{cc}      
 a&0\\      
 c&1\\      
 \end{array}      
 \hspace{0.0in} \right) \;\; ,      
\end{equation}      
where
\begin{equation}
c=\sqrt{\frac{(1-\alpha)(1-\beta)}{ \alpha \beta }}
\quad , \quad  a
=\sqrt{\frac{1}{\alpha \beta}}\quad .
\end{equation}

In order to compute expectation values, it is convenient
to diagonalize  $K$ (see (\ref{Kdef})). It turns out that this can be done for
$\alpha\ne\beta$; we discuss this case first, but omit details of
the diagonalization. However, note that the eigenvalues of $K$ are
$1/\alpha$ and $1/\beta$.

Using the diagonalization, it is straightforward to compute 
$z_N=\langle W_1|K^N|V_1\rangle$   
by writing $|V_1\rangle$ as a superposition of eigenvectors of $K$
and one obtains
\begin{equation}
z_N=\frac{(1-\beta)\alpha\beta^{-N}-(1-\alpha)\beta\alpha^{-N}}{\alpha-\beta}
\quad.
\label{zp=1a}
\end{equation}
Equation (\ref{Z2}) then leads to
\begin{equation}
Z_N=\frac{(1-\beta^2)\alpha\ \beta^{-N}
-(1-\alpha^2) \beta\ \alpha^{-N}}{\alpha-\beta}\quad.
\end{equation}
Thus, the current $J_N$ follows from (\ref{Jeq}):
\begin{equation}
J_N=\alpha\beta\frac{(1-\beta)\alpha^N-(1-\alpha)\beta^N}
{(1-\beta^2)
\alpha^{N+1}-(1-\alpha^2)\beta^{N+1}}\quad.
\end{equation}
 From this equation, it is clear that there are only two
phases (for $\alpha\ne\beta$), a high density phase when
$\alpha>\beta$ and a low density
phase when $\beta>\alpha$. As $N\rightarrow \infty$, $J_N$ 
approaches $\beta/(1+\beta)$ or
$\alpha/(1+\alpha)$, respectively. These results were conjectured in \cite{RSSS,TE}. 

With the diagonalization and (\ref{density}) it is straightforward 
to work out a formula for the density profile 
$\langle\tau_i\rangle_N(\alpha,\beta)$ ($i=1,...,N$):
\begin{equation}\label{densityp=1}
\langle\tau_i\rangle_N=\frac{\alpha^{N+1}(1-\beta)-\beta^{N+1}\alpha(1-\alpha)-
(\beta/\alpha)^i (1-\alpha)(1-\beta)\alpha^{N+1}}
{(1-\beta^2)\alpha^{N+1}-(1-\alpha^2)\beta^{N+1}}\, .
\end{equation}
It follows that in the limit $N\rightarrow\infty$ 
the density profile in the high density phase $\alpha>\beta$ is given
by
\begin{equation}
\langle\tau_i\rangle=\frac{1}{1+\beta} - \frac{1-\alpha}{1+\beta}
(\beta/\alpha)^{i}\, .
\end{equation}
\noindent
This implies that the density is constant for $i\gg 1$ (in particular,
in the bulk and near the right boundary), and decays
exponentially near the left boundary (on a scale given by $\log{(\alpha/\beta)}$)
towards the bulk density:
\begin{equation}
\langle\tau\rangle_{\rm bulk}=\frac{1}{1+\beta}\quad.
\end{equation}
The corresponding formulae for the
the low density phase $\beta>\alpha$ can be obtained easily directly from
(\ref{densityp=1}) or via the particle hole symmetry (\ref{symm}). One obtains
\begin{equation}
\langle\tau\rangle_{\rm bulk}=\frac{\alpha}{1+\alpha} \quad,
\end{equation}
and an exponentially growing density profile near the right boundary.
These values for the bulk densities are in agreement with 
conjectures in \cite{RSSS,TE}.

The two-point correlation function (for $\alpha\ne \beta$) can be
computed via (\ref{twopoint}) and is given by
the following expression:
\begin{eqnarray}\label{finitecorr}
\langle\tau_i\tau_j\rangle&=&f(\alpha,\beta,N)^{-1}\times
\left\{ \ 
\beta^{N+1}\alpha^2(\alpha-1)(1+\beta)
+\alpha^{N+1}(1+\alpha)(1-\beta)\right. \\\nonumber 
&&\hspace{0.8in}+\beta^{N+1}\alpha(\alpha-1)(1+\beta)(-\alpha)^{j-i}
+\alpha^{N+1}(1+\alpha)(\beta-1)(-\beta)^{1+j-i}\\\nonumber
&&\hspace{0.8in}+\alpha^{N+1}(\alpha-1)(1+\alpha)(1-\beta)(\beta/\alpha)^i
\\\nonumber
&&\hspace{0.8in}+\alpha^{N+1}(\alpha-1)(\beta-1)(\alpha-\beta)
(-\beta)^j(-\alpha)^{-i}\\\nonumber
&&\left. \hspace{0.8in}+\alpha^{N+1}(\alpha-1)(1-\beta)(1+\beta)\alpha
(\beta/\alpha)^j\ \right\}\quad,
\end{eqnarray}
with $i\le j$ and
\begin{equation}
f(\alpha,\beta,N)=(1+\alpha)(1+\beta)
\left[ \ (1-\beta^2)\alpha^{N+1}-(1-\alpha^2)\beta^{N+1}\ \right]\quad .
\end{equation}
 From this one can derive asymptotic expressions; for example,
in the high density phase ($\alpha > \beta$) the bulk two-point correlation
function ($j=i+r$, $1<<i\le i+r$) is
\begin{equation}\label{corrlim}
\langle\tau_i\tau_{i+r}\rangle_{\rm bulk}=\frac{1+\beta(-\beta)^r}{(1+\beta)^2}\quad.
\end{equation}
Note the oscillating nature of (\ref{corrlim}), which can be interpreted 
as a particle-hole attraction which is created by the simultaneous
updating (see introduction). The corresponding truncated correlation function,
\begin{equation}
g(i,j)=\langle\tau_i\tau_j\rangle - \langle\tau_i\rangle\langle\tau_j\rangle
=\frac{\beta(-\beta)^{j-i}}{(1+\beta)^2}\, ,
\end{equation}
decays for $\beta\ne 1$ exponentially to zero on a scale $1/{|\ln \beta|}$. 

Let us now turn to the case {$\alpha=\beta$}.
In that case, $K$ cannot be diagonalized, but there exists
a similarity transformation which reduces $K$ to Jordan
normal form and makes all the necessary 
calculations straightforward. We just list the results:
\begin{equation}
z_N=\frac{1}{\alpha^N}\left[\, N(1-\alpha)+1\, \right] \quad,
\label{zp=1b}
\end{equation}
\begin{equation}
Z_N=\frac{1}{\alpha^N}\left[\, N(1-\alpha^2)+\alpha^2+1\, \right]
\quad,
\end{equation}
\begin{equation}
J_N=\alpha\frac{N(1-\alpha)+\alpha}{N(1-\alpha^2)+1+\alpha^2}\quad ,
\end{equation}
and          
\begin{equation}\label{finitedens}
\langle\tau_i\rangle_N=\frac{(1-\alpha)^2\,i+\alpha N(1-\alpha)+\alpha}{N
(1-\alpha^2)+1+\alpha^2}\, ,
\end{equation}
which yields for large $N$ 
\begin{equation}\label{asympdens}
\langle\tau_i\rangle_N = \frac{\alpha+(1-\alpha)\,(i/N)}{1+\alpha}
+O(1/N) \quad.
\end{equation}
 The error term is small when $N(1-\alpha)\gg 1$.
Again, these expressions coincide with the expected formulae \cite{RSSS}.
The linear profile in (\ref{asympdens}) can be interpreted as arising
from a uniform superposition of states with localized shocks.

The two point correlation function is now
\begin{eqnarray}\label{corr}
\langle\tau_i\tau_j\rangle=t(\alpha,N)^{-1}\times \Bigl\{
(1-\alpha)(1-\alpha^2)i+\alpha(1-\alpha)(1-\alpha^2)j+N\alpha^2(1-\alpha^2)
+2\alpha^2\nonumber\\
+(-\alpha)^{(j-i)}\Bigl( \, \alpha(1-\alpha^2)(i-j)+N\alpha(1-\alpha^2)+
\alpha(1+\alpha^2)\, \Bigr) \Bigr\}
\end{eqnarray}
where $i,j=1,2,\ldots,L$, $i\le j$ and
$t(\alpha,N)=(1+\alpha)^2\left[ \, N(1-\alpha^2)+1+\alpha^2\, \right]$. 
When $\alpha=1$ this reduces to
\begin{equation}\label{prim}
\langle \tau_i\tau_j\rangle _{\alpha=1}=\frac{1}{4}
\left[1+(-1)^{(j-i)}\right]\, ,
\end{equation}
which is the expected result since then in the steady state
only the two configurations in which there are no particle-particle
or hole-hole pairs occur with non zero probability. The 
alternating structure of (\ref{prim}) does not show up in
the density profile (which is flat here) because these two
configurations have equal weights.

For $\alpha\ne1$, the truncated correlation function $g(i,j)$ simplifies for
large $N$: 
 \begin{equation}
 g(i,j) = \frac{1}{(1+\alpha)^2}\Bigl\{(i/N)(1-(j/N))(1-\alpha)^2
          +(-\alpha)^r\alpha(1+(r/N))\Bigr\} + O(1/N)\,,
 \end{equation} 
 where $r=j-i$.  Let us briefly discuss the behavior of $g(i,j)$ for fixed
$i$ and $j=i,i+1,\ldots,N$.  The oscillating part of $g(i,j)$, which is not
present with other updating schemes, decays exponentially with $j$ on a scale
$1/{|\ln\alpha}|$.  Therefore, for sufficiently large $j$, $g(i,j)$ decays
linearly to zero with slope $-(\frac{1-\alpha}{1+\alpha})^2\frac{i}{N^2}$. 
Figure 1 shows two examples of $g(i,j)$ for a system of $100$ sites.  The
strong oscillations present for $\alpha=\beta=0.9$ arise because the density
in the system is nearly at its maximum value of $1/2$, so that if a site is
empty its two nearest neighbors are probably occupied.  When
$\alpha=\beta=0.1$, on the other hand, each site is for some typical
configurations in a region of low density (if the shock is to its right) and
for some in a region of high density (if the shock is to its left), so that
the truncated correlations are positive. 

We now turn to the calculation of 
the fluctuations in the number $M$ of particles
in the system, still considering the case $\alpha=\beta$. We write 
\begin{eqnarray}
\langle M\rangle &=&\sum_{i=1}^{N}\langle \tau_i\rangle \\
\Delta^2&=&\langle M^2\rangle -\langle M\rangle ^2=
2\sum_{i<j}^{N}\langle \tau_i\tau_j\rangle +\langle M\rangle -{\langle M\rangle }^2
\end{eqnarray}
and must sum up the expressions
(\ref{finitedens}) and (\ref{corr}), respectively. The first summation
is trivial:
\begin{equation}
\langle M\rangle =\frac{N^2(1-\alpha^2)+
N(1+\alpha^2)}{2N(1-\alpha^2)+2(1+\alpha^2)}\quad.
\end{equation}
For $\alpha=\beta=1$ we obtain the expected result $\langle M\rangle _{\alpha=1}=N/2$.
The summation of (\ref{corr}) is more tedious. It is convenient to use
$\sum_{j,i<j}^{N}h(r)=\sum_{r=1}^{N-1}(N-r)h(r)$, where $h(r)$ is an
arbitrary function of $r$. One is then left with well-known sums of the form
$\sum_{r=1}^{N-1}r^k(-\alpha)^r$
($k=0,1,2$). Altogether one obtains
\newpage\begin{eqnarray}\label{worm}
\nonumber\Delta^2(\alpha,N)&=&
t(\alpha,N)^{-1}\times \Bigl\{
\frac{N^3}{3}[1+\alpha-\alpha^3-\alpha^4]-N^2\alpha^2[\frac{1-3\alpha-\alpha^2
-\alpha^3}{1+\alpha}]\nonumber\\
&-&2N\alpha^2[\frac{2+\alpha+\alpha^2}{1+\alpha}]+
(1-\alpha+\alpha^2)(\frac{2\alpha}{1+\alpha})^2\nonumber\\
&-&(-\alpha)^N(\frac{2\alpha}{1+\alpha})^2[N(1-\alpha^2)+
1-\alpha+\alpha^2]\Bigr\}\nonumber\\
&+&\langle M\rangle -\langle M\rangle ^2\quad.
\end{eqnarray}
For $\alpha=1$ (\ref{worm}) yields $1/4$\ for $N$ odd, and $0$ for
$N$ even, as expected.

We now take $N\gg 1$ and keep only the highest order in $N$. This gives
\begin{equation}
\Delta^2=\frac{N^2}{12}\frac{(1-\alpha)^2}{(1+\alpha)^2}\quad,
\end{equation}
which can be rewritten, using (\ref{asympdens}), as
\begin{equation}\label{numberfluc}
\Delta^2=\frac{N^2}{12}(\rho_{\rm right}-\rho_{\rm left})^2\quad,
\end{equation}
where $\rho_{\rm right}$($\rho_{\rm left}$) is the asymptotic
density at position $N$($1$) given by (\ref{asympdens}).
This is precisely the result which is to be expected when one
considers the linear profile as a superposition of uniformly
distributed random shock positions (step functions).
\section{Derivation of the phase diagram for general $p$} 
\label{phasediagram}
\setcounter{equation}{0} 

In this section we determine the asymptotic behavior, 
for all values of $\alpha, \beta, p$, of the quantity
$z_N=Ewv{K^N}$ introduced in (\ref{littlez}) and hence, through
(\ref{Jeq}), of the current $J_N$; the different possible asymptotic forms
determine the distinct phases of the model.  Our method is to study the
 generating function 
\begin{equation}
\Theta_0(\lambda)
   \equiv \sum_{N=0}^\infty\lambda^N z_N.\label{Thetadef}
\end{equation}
We will use the explicit representation (\ref{W1V1})--(\ref{E1}) of the
operators $D_1$ and $E_1$ and the vectors $\Ws$ and $\Vs$, and will write
$\rv{n}$, $n=0,1,\ldots$, for the basis of the space on which $D_1$ and
$E_1$ act, and $\lv{n}$ for the dual basis, so that $\Vs=\rv{0}$ and
$\Ws=\lv{0}$. Note that  $z_0=\langle W_1|V_1\rangle=1$.

From (\ref{D1}) and (\ref{E1}) it follows that $K\equiv E_1+D_1+p$ has the form 
 \begin{equation}
{K} = \left( \begin{array}{llllll}      
 c&b&0&0&. &\hspace{0.2in}.\\      
 b&2-p&(1-p)^{1/2}&0&& \\      
 0&(1-p)^{1/2}&2-p&(1-p)^{1/2}&&\\      
 0&0&(1-p)^{1/2}&2-p&. &\\      
 . &&&&. &\hspace{0.2in}. \\      
 . &&&&&\hspace{0.2in}.      
 \end{array}\right)\;, \label{K1}
 \end{equation}
where $c=p(\alpha+\beta-\alpha\beta)/\alpha\beta$
and $b$ is given by (\ref{bsq}).
From (\ref{K1}) we find
 \begin{eqnarray} 
\langle 0 |K^{N+1} |0\rangle
&=& c \langle 0 |K^{N} |0\rangle + b \langle 1 |K^{N} |0\rangle\,,
\label{Krecur1}\\
\langle 1 |K^{N+1} |0\rangle
&=& b \langle 0 |K^{N} |0\rangle + (2-p) \langle 1 |K^{N} |0\rangle
+ (1-p)^{1/2} \langle 2 |K^{N} |0\rangle\,,
\label{Krecur2}
\end{eqnarray}
and for $n>1$,
\begin{eqnarray}
\hspace*{-1.2cm}\langle n |K^{N+1} |0\rangle
&=& (1-p)^{1/2} \langle n-1 |K^{N} |0\rangle
\nonumber \\
&&+ (2-p) \langle n |K^{N} |0\rangle
    + (1-p)^{1/2} \langle n+1 |K^{N} |0\rangle\,.
\label{Krecur3}
\end{eqnarray}
If we now define the generating functions
\begin{equation}
\Theta_n=\sum_{N=0}^{\infty}
\lambda^N \langle n |K^{N} |0\rangle\,,
\end{equation}
 we easily obtain from (\ref{Krecur1})--(\ref{Krecur3}),
using $\langle n | 0 \rangle=\delta_{n,0}$,  that
\begin{eqnarray}
(1-c \lambda) \Theta_0
&=&   b \lambda \Theta_1+1\,, \label{Lrecur1}\\
(1- (2-p) \lambda) \Theta_1
&=& b \lambda \Theta_0  + (1-p)^{1/2} \lambda \Theta_2\,, \label{Lrecur2}
\end{eqnarray}
and for $n>1$,
\begin{equation}
\hspace*{2.4cm}
(1- (2-p) \lambda) \Theta_n
=  (1-p)^{1/2}\lambda \Theta_{n-1} + (1-p)^{1/2} \lambda  \Theta_{n+1}\,.
\label{Lrecur3}
\end{equation}
The  solution of (\ref{Lrecur3}) is 
 \begin{equation}
\Theta_n = A u^n\quad \mbox{for} \quad n >0\,,
\label{Lnsol}
\end{equation}
where $A$ is a constant to be determined
from (\ref{Lrecur2}) and
\begin{equation}
u=    \frac{1-\lambda(2-p)-\sqrt{(1+\lambda p)^2-4\lambda}   
             }{
               2\lambda (1-p)^{1/2} }\,.  \label{udef}
\end{equation}
(the positive root is discarded being non-analytic at
$\lambda =0$).
Writing $\Theta_1=Au$ and $\Theta_2=Au^2$ in (\ref{Lrecur1}) and
(\ref{Lrecur2}), and eliminating $A$ from the resulting equations,
yields
 \begin{equation}
\Theta_0 =\frac{1}{1-c\lambda - b^2 \lambda u/(1-p)^{1/2}}
\label{L0}
\end{equation}
 From (\ref{L0}), (\ref{udef}), and some tedious algebra, we find that
 \begin{equation}
   \Theta_0(\lambda)=\frac{
 \alpha\beta \left[2(1-p)(\alpha\beta-p^2\lambda)
    -\alpha\beta b^2(1-p\lambda)
    -\alpha\beta b^2\sqrt{(1+p\lambda)^2-4\lambda}\,\right]}
 {2p^4(1-\beta)(1-\alpha)(\lambda-\lhd)(\lambda-\lld)},
 \label{Thfinal}
 \end{equation}
 where
 \begin{eqnarray}
 \lambda_{\rm ld} &=& \frac{\alpha(p-\alpha)}{ p^2(1-\alpha)}\,,\\
 \lambda_{\rm hd} &=& \frac{\beta(p-\beta)}{ p^2(1-\beta)}\,.
 \end{eqnarray}

 Equation \(Thfinal) shows that $\Theta_0$ has square root singularities at the
two points
 \begin{equation}
 \lambda_{\rm mc}^\pm = \frac{2-p\pm 2\sqrt{1-p}}{ p^2} \label{lammc}
 \end{equation}
which, if we assume that the parameters $\alpha$, $\beta$, and $p$ lie
in the relevant range $0\le \alpha,\beta,p\le1$, are 
on the positive real axis.  Thus $\Theta_0$ is naturally double valued and has
single valued determinations (branches) on each of two {\it sheets}---copies
of the complex plane cut along the real axis between the two roots.  We are
primarily interested in the behavior on the {\it first sheet}, the plane on
which, for $\lambda$ small and real, the square root in \(Thfinal) is
positive and hence $\Theta_0(0)=1$ (see (\ref{Thetadef})).  $\Theta_0$ also has
two simple poles, at $\lld$ and $\lhd$. 
 
As discussed in Appendix~\ref{appendixa}, the coefficients $z_N$ in the power
series \(Thetadef) will grow as $N^\gamma\lambda_0^N$, where $\lambda_0$ is
the singularity of $\Theta_0(\lambda)$ on the first sheet nearest to the origin
($\lambda_0>1$ always) and $\gamma$ is determined by the nature of that
singularity.  Thus
 \begin{equation}
  \lim_{N\to\infty}\frac{z_N}{z_{N+1}} = \lambda_0, \label{limrat}
 \end{equation}
  and hence
 \begin{equation}
  \lim_{N\to\infty}J_N=\frac{p\lambda_0}{1+p\lambda_0}. \label{cur2}
 \end{equation}
Three regions in the parameter space must be considered according
to which of the three singularities is closest to the origin.
As we shall see, the singularities $\lld$ and $\lhd$
may or may not be present on the first sheet of the complex plane. 
For the parameter values where one or both do occur they
are closer to the origin than  $\lambda_{\rm mc}^-$.
 It is convenient to introduce the quantity 
\begin{equation}
q=q(p)=1-\sqrt{1-p}
\end{equation}
in discussing the resulting phase diagram (thus $2q-p=q^2$ and
$\lmc^-=(q/p)^2$). The phase diagram is shown in figure 2.

  \smallskip\noindent
 (i) {\bf The maximum current region:} $q<\alpha$ and $q<\beta$ 
(region C in figure~2).

 For these parameter values the numerator in \(Thfinal) vanishes at
$\lambda_{\rm ld}$ and $\lambda_{\rm hd}$ when the square root is positive:
the poles lie on the second sheet and the singularity closest to the origin
in the first sheet is $\lambda_{\rm mc}^{-}$.  Then \(alg) implies that
 \begin{equation}
  z_N = \frac{\alpha^2\beta^2b^2(1-p)^{1/4}\sqrt{\lmc^-}}
   {2\sqrt\pi (\alpha-q)^2(\beta-q)^2}
 \frac{1}{N\sqrt{N}(\lmc^-)^N} + O(N^{-5/2}(\lmc^-)^N),
 \label{znmc}
 \end{equation}
 and hence from \(cur2),
 \begin{equation}
  \lim_{N\to\infty}J_N=\frac{p\lambda_{\rm mc}^{-}}{1+p\lambda_{\rm mc}^{-}}
   = \frac{1-\sqrt{1-p}}{2}\,.
 \end{equation}
 Note that the prefactor in (\ref{znmc}) is singular at the boundaries
$\alpha=q$, $\beta=q$ of the maximum current region; near these boundaries
one needs larger values of $N$ for the leading term in (\ref{znmc}) to well
approximate $z_n$. 

\smallskip\noindent
 (ii) {\bf The low density  region:} $\alpha<\beta$ and $\alpha<q$
(region A in figure~2).

 In this region the pole $\lambda_{\rm ld}$ lies on the first sheet and is
in fact the singularity of $\Theta_0(\lambda)$ closest to the origin, and  from
\(simple),
 \begin{equation}
  z_N = \frac{\beta (p+\alpha^2-2\alpha)}
   {(p-\alpha)(\beta-\alpha)}
 \frac{1}{(\lld)^{-N}} + o(s^{-N}),
 \label{znld}
 \end{equation}
 for some $s>\lld$.   Thus from \(cur2),
 \begin{equation}
  \lim_{N\to\infty}J_N=\frac{p\lambda_{\rm ld}}{1+p\lambda_{\rm ld}}
   = \frac{\alpha(p-\alpha)}{ p-\alpha^2}\,.
 \end{equation}

\smallskip\noindent
 (iii) {\bf The high density  region:} $\beta<\alpha$ and $\beta<q$
(region B in figure~2).

 The asymptotic behavior of $z_N$ here is obtained from \(znld) by
interchanging $\alpha$ and $\beta$ and replacing $\lld$ by $\lhd$. In
particular, 
 \begin{equation}
  \lim_{N\to\infty}J_N
   = \frac{\beta(p-\beta)}{ p-\beta^2}\,.
 \end{equation}
 %
\section {Asymptotics of the density for general $p$}
\label{asymptotics}
\setcounter{equation}{0}

In this section we calculate the behavior of the particle density near the
left end of the system in the limit of infinite system size
and for all values of $\alpha$, $\beta$ and $p$; behavior near
the right end can be recovered from the symmetry (\ref{symm}).  For
$m,n\ge0$ let
 \begin{equation}
t_{m,n}=\lim_{N\to\infty}\frac{Ewv{K^{N-m-n}D_1^mK^n}}{z_N}
 \end{equation}
 and introduce the generating function
 \begin{equation}
\Psi(x,y)=\sum_{m,n\ge0}x^my^nt_{m,n}\,.
 \end{equation}
 Our goal is to calculate $\Psi_x(0,y)$.  For it follows from
(\ref{density}) and \(limrat) that $\rho_n$, the density at the $(n+1)^{\rm
st}$ site to the left of the right boundary in the infinite volume limit
(where $n=0,1,\ldots$), is given by
 \begin{equation} \rho_n = \lim_{N\to\infty}\langle\tau_{N-n}\rangle_N
   = \lim_{N\to\infty}\frac{Ewv{K^{N-1-n}(D_1+p)K^n}}
      {z_N+pz_{N-1}}
   = \frac{t_{1,n}+p\lambda_0}{1+p\lambda_0}, 
\end{equation}
 so that the generating function $\Phi(y)$ for the $\rho_n$ is
 \begin{equation}
  \Phi(y)=\sum_{n\ge0}y^n \rho_n
    =\frac{1}{1+p\lambda_0}\sum_{n\ge0}y^n(t_{1,n}+p\lambda_0)
    = \frac{1}{1+p\lambda_0}\left(\Psi_x(0,y)+\frac{p\lambda_0}{1-y}\right).
 \label{chi}
 \end{equation}

Now $t_{0,n} = 1$ for all $n$, so that $\Psi(0,y)=(1-y)^{-1}$, and from
(\ref{D1V1}) and \(limrat),
$t_{m,0} = \bigl(\lambda_0p(1-\beta)/\beta\bigr)^m$ for all $m$, so that
$\Psi(x,0)=\beta/(\beta - x\lambda_0p(1-\beta))$.  From (\ref{D1E1con}) it
follows that 
$D_1K=D_1^2+(1-p)K+pD_1$ and this, together with \(limrat),
implies that for $m,n\ge1$, $t_{m,n}$ satisfies the recursion
 \begin{equation}
t_{m,n} = t_{m+1,n-1}+\lambda_0(1-p)t_{m-1,n}+\lambda_0pt_{m,n-1}. 
 \end{equation}
 Thus
 \begin{eqnarray}
\Psi(x,y)&=& \sum_{m\ge0}x^mt_{m,0} + \sum_{n\ge1}y^nt_{0,n}
     + \sum_{m,n\ge1}x^my^nt_{m,n} \nonumber \\
   &=& \Psi(x,0)+[\Psi(0,y)-1]
+\sum_{m,n\ge1}x^my^n[t_{m+1,n-1}+\lambda_0(1-p)t_{m-1,n}+\lambda_0pt_{m,n-1}]
    \nonumber \\
   &=& \Psi(x,0)+[\Psi(0,y)-1]
        +\frac{y}{ x} [\Psi(x,y)-\Psi(0,y)-x\Psi_x(0,y)] \nonumber\\
   &&\hskip50pt  +\lambda_0(1-p)x[\Psi(x,y)-\Psi(x,0)]
     +\lambda_0py [\Psi(x,y)-\Psi(0,y)]. \label{Psi}
 \end{eqnarray}
 Multiplying \(Psi) by $-x$, collecting terms, and using the relation
$\Psi(0,y)-1=y\Psi(0,y)$, we see that $\Psi$ satisfies the equation
 \begin{equation}
   D(x,y)\Psi(x,y) = A(x,y) + xy\Psi_x(0,y), \label{Psi2}
 \end{equation}
 where
 \begin{eqnarray}
  D(x,y) &=& \lambda_0(1-p)x^2-(1-\lambda_0py)x+y,\\
  A(x,y) &=&  -y[x(1-\lambda_0p)-1]\Psi(0,y) -x[1-\lambda_0(1-p)x]\Psi(x,0).
 \end{eqnarray}

 Now the branch of the curve $D(x,y)=0$ given by $x=\xi_-(y)$, where 
 \begin{equation}
  \xi_\pm (y)
 =\frac{1-\lambda_0py \pm \sqrt{\Delta(y)}}{2\lambda_0(1-p)}
 \end{equation}
 with
 \begin{equation}
	\label{deltay}
 \Delta(y) = (1+\lambda_0py)^2-4\lambda_0y,
 \end{equation}
 are the roots in $x$ of $D(x,y)=0$, passes through the origin.  But
$\Psi(x,y)$ is analytic at the origin, so that \(Psi2) can hold only if the
right hand side vanishes on this curve.  This yields the desired equation for
$\Psi_x(0,y)$:
 \begin{equation}
   \Psi_x(0,y) = -\frac{A(\xi_-(y),y) }{ \xi_-(y)y}
  = \frac{\xi_-(y)(1-\lambda_0p)-1}{\xi_-(y)(1-y)}
     +\frac{\beta[1-\lambda_0(1-p)\xi_-(y)]}{
            y[\beta- \xi_-(y)\lambda_0p(1-\beta)]}.\label{psi1}
 \end{equation}
 If we insert (\ref{psi1}) into from \(chi) and rationalize the resulting
expression we obtain
 \begin{eqnarray}
  \Phi(y) &=& \frac{1}{1+p\lambda_0}
\left(\frac{2y-1+\lambda_0py}{2y(1-y)}\ 
    -\ \frac{\sqrt{\Delta(y)}}{2y(1-y)}\right.
\nonumber\\
   &&\hskip-10pt+\ \left.
     \frac{\beta(p-\beta)\sqrt{\Delta(y)}}{
    2yp^2(1-\beta)(\lhd-\lambda_0y)}
   \ +\ \frac{\beta((p-\beta)-\lambda_0py(2-\beta-p))}{
    2yp^2(1-\beta)(\lhd-\lambda_0y)}
  \right). \label{chi2}
 \end{eqnarray}
 We will always assume that the parameters in \(chi2) lie in the physical
region $0\le\alpha,\beta,p\le1$.  Under this assumption the two roots of the
equation $\Delta(y)=0$ lie on the positive real axis, so that we may regard
$\Phi$ as defined on two sheets, as we did $\Theta_0$ in the previous section. 
The first sheet corresponds, for $y$ real and small, to $\sqrt{\Delta(y)}>0$
in \(chi2). 

 From \(chi2) we see that the singular points of $\Phi$ (which may coincide
for some parameter values) are:\\
 \hbox to 20pt{\hss$\bullet$\hss} A simple pole at $y=1$.\\
 \hbox to 20pt{\hss$\bullet$\hss} Two square root singularities $y_\pm $, the
roots of the equation $\Delta(y)=0$; from (\ref{lammc}),
 \begin{equation}
  y_\pm =\lmc^\pm /\lambda_0.\label{ypm}
 \end{equation}
 Since $\lambda_0\le\lmc^-$, these singularities satisfy
$1\le y_-\le y_+$.\\
 \hbox to 20pt{\hss$\bullet$\hss}  An apparent simple pole at 
 \begin{equation}
    y_1=\frac{\lambda_{\rm hd}}{\lambda_0}
 = \frac{\beta(p-\beta)}{\lambda_0 p^2(1-\beta)}.\label{y1}
 \end{equation}
 However, the numerators of the third and fourth terms in \(chi2) may be equal
in magnitude and opposite in sign when $y=y_1$, cancelling this singularity;
from \(y1) and \(chi2) we find that this happens when
 \begin{equation}
    p(1-\beta)-\beta(2-\beta-p)  = - p(1-\beta)\sqrt{\Delta(y_1)}.
   \label{cancel}
 \end{equation}
 A little algebra shows that the squares of the two sides in \(cancel) are
equal, so that \(cancel) holds on the first sheet, and the pole at $y_1$ is
absent there, if $p(1-\beta)-\beta(2-\beta-p)\le0$, i.e., if
$\beta \ge q$. \\
 \hbox to 20pt{\hss$\bullet$\hss}  
 An apparent simple pole at $y=0$.  From (\ref{chi}), however, it follows
that $\Phi$ is regular at the origin on the first sheet.  This can also be
seen directly from \(chi2) using $\sqrt{\Delta(0)}=1$.  

We now analyze this generating function in the various regions of the phase
plane of the system.

 \smallskip\noindent
 (i) {\bf The maximum current region:} $q<\alpha$ and $q<\beta$ 
(region C in figure 2).

\nopagebreak
In this region $\lambda_0=\lmc^-$, so that from \(ypm) the square root 
singularity $y_-$ coincides with the pole at $y=1$; thus $\Delta(y)$
has a factor $(1-y)$ and from (\ref{deltay}),
 \begin{equation}
  \Delta(y)=(1-y)(1-p^2(\lmc^-)^2y).\label{newDel}
 \end{equation}
  Since $q<\beta$, there is no pole at $y_1$ on the first sheet, and
thus $y=1$ is the singularity closest to the origin and controls the
asymptotics of the coefficients $\rho_n$ of $\Phi$.  We write 
$\rho_n=\rho_n^{(1)}+\rho_n^{(2)}+\rho_n^{(3)}$, where $\rho_n^{(i)}$ is
the contribution from the $i^{\rm th}$ term in \(chi2) (the fourth term is
regular at $y=1$), and calculate
the asymptotic form of each $\rho_n^{(i)}$ in turn.   

The first term in \(chi2) is $f_1(y)(1-y)^{-1}$, where
$f_1(y)=(2y-1+\lmc^-py)\Big/\Bigl(2y(1+p\lmc^-)\Bigr)$. Thus, 
from \(simple) and \(mone), 
 \begin{equation}
 \rho_n^{(1)} = f_1(1) + o(s^{-n}) = \frac{1}{2}+ o(s^{-n}) \label{rho1}
 \end{equation}
 for any $s>1$.  The second term is $-f_2(y)(1-y)^{-1/2}$, where
$f_2(y)=\sqrt{1-p^2(\lmc^-)^2y}\Big/\Bigl(2y(1+p\lmc^-)\Bigr)$ and we 
have used \(newDel).  
From \(alg), \(mhalf), and \(half), then,
 \begin{eqnarray}
 \rho_n^{(2)} &=& -f_2(1)\left(1-\frac{1}{8n}\right)\frac{1}{\sqrt{\pi n}} 
    - f_2'(1)\frac{1}{2n\sqrt{\pi n}}\ +\  O(n^{-5/2})\label{rho2}\\
  &=& -\frac{\sqrt{1-q}}{2}\;\frac{1}{\sqrt{\pi n}} 
  \ +\ \frac{p^2+6q^2
      (1-q)}{32q^2\sqrt{1-q}}\;
   \frac{1}{ n\sqrt{\pi n}}\ +\  O(n^{-5/2}).\nonumber
 \end{eqnarray}
 The third term is $f_3(y)(1-y)^{1/2}$, where 
 \begin{equation}
f_3(y)=\frac{\beta(p-\beta)\sqrt{1-p^2(\lmc^-)^2y}}{
        2yp^2(1-\beta)(1+p\lmc^-)(\lhd-\lmc^-y)}.
 \end{equation}
As above
 \begin{eqnarray}
 \rho_n^{(3)} &=& -f_3(1)\frac{1}{2n\sqrt{\pi n}}\ +\  O(n^{-5/2})\nonumber\\
  &=& \frac{\sqrt{1-q}\beta(p-\beta)}{4(\beta-q)^2}
   \frac{1}{ n\sqrt{\pi n}}\ +\  O(n^{-5/2}).\label{rho3}
 \end{eqnarray}
 Adding \(rho1), \(rho2), and \(rho3) gives the density $\rho_n$ to order
$n^{-3/2}$:
 \begin{eqnarray}
 \label{asymc}
 \rho_n &=& \frac{1}{2}\ -\ \frac{\sqrt{1-q}}{2}\;\frac{1}{\sqrt{\pi n}}\label{rhomc}\\ 
  &&\hskip20pt
  \ +\ \left(
    \frac{p^2+6q^2(1-q)}{32q^2\sqrt{1-q}}
  \ + \ \frac{\beta\sqrt{1-q}(p-\beta)}{ 4(\beta-q)^2}\right)
   \frac{1}{ n\sqrt{\pi n}}\ +\  O(n^{-5/2}).\nonumber
 \end{eqnarray}
   In the $p\searrow0$ limit, with the scaling $\alpha=p\bar\alpha$,
$\beta=p\bar\beta$, this result corresponds to that determined in
\cite{DEHP}.  The last coefficient (\ref{asymc}) is singular on the
$\beta=q$ boundary of the maximum current region and in particular 
at $p=1$; see the comment following (\ref{znmc}). 

 \smallskip\noindent
 (ii) {\bf The low density  region:} $\alpha<\beta$ and $\alpha<q$
(region A in figure 2).

  Here $\lambda_0=\lld$.  Since $y_\pm =\lmc^\pm /\lambda_0>1$, the square root
singularities lie strictly to the right of the pole at $y=1$.  Let us again
write $\rho_n=\rho_n^{(1)}+\rho_n^{(2)}+\rho_n^{(3)}+\rho_n^{(4)}$, with
$\rho_n^{(i)}$ the contribution from the $i^{\rm th}$ term in \(chi2).
Each of the first two terms in \(chi2) has a simple pole at $y=1$, so that
from \(alg),
 \begin{equation}
\rho_n^{(1)}+\rho_n^{(2)}=
  \left[\frac{2y-1+\lld py- \sqrt{\Delta(y)}}{
      2y(1+p\lld)}\right]_{y=1}+o(y_-^{-n})
   =\frac{\alpha(1-\alpha)}{ p-\alpha^2}+o(y_-^{-n}),  \label{ldr12}
 \end{equation}
 where we have used
      $\sqrt{\Delta(1)}=(p-2\alpha+\alpha^2)/(p(1-\alpha))$ 
(note that this is
the positive square root).  To go further in the asymptotics we must consider
separately two subregions of the low density region, and their common
boundary (see Figure~2). 
 
 \smallskip\noindent
 {\bf Subregion A$\,$I:} $\beta<q$.

In this subregion $\lld<\lhd<\lmc^-$.  Since $\beta<q$, the
pole at $y_1=\lhd/\lld>1$ lies on the first sheet and satisfies
$1<y_1<y_-$, and thus makes the next contribution to the asymptotics beyond
\(ldr12). Thus
 \begin{eqnarray}
  \rho_n^{(3)}+\rho_n^{(4)} &=&
     \left[\frac{\beta\left((p-\beta)\sqrt{\Delta(y)}
    +(p-\beta)-\lld py(2-\beta-p)\right)}{
    2yp^2(1+p\lambda_0)(1-\beta)\lhd}\right]_{y=y_1}
  \left(\frac{\lhd}{\lld}\right)^{-n}+o(s^{-n})
   \nonumber\\
 &=&\frac{(1-\alpha)(p-2\beta+\beta^2)}{(p-\alpha^2)(1-\beta)}
  \left(\frac{\alpha(p-\alpha)(1-\beta)}{\beta(p-\beta)(1-\alpha)}\right)^{n+1}
   +o(s^{-n}),\label{ldr34}
 \end{eqnarray}
 for some $s>\lhd/\lld$.  The asymptotics to order $o(s^{-n})$ are obtained by
adding \(ldr12) and \(ldr34):
 \begin{equation}
  \rho_n\ =\ \frac{\alpha(1-\alpha)}{ p-\alpha^2} \ + \
   \frac{(1-\alpha)(p-2\beta+\beta^2)}{(p-\alpha^2)(1-\beta)}
  \left(\frac{\alpha(p-\alpha)(1-\beta)}{\beta(p-\beta)(1-\alpha)}\right)^{n+1}
   \ +\ o(y_-^{-n}).\label{ldrI}
 \end{equation}
   Further corrections, which arise from the singularity at $y_-$,
could be calculated; the leading order is $O(y_-^{-n}/n^{3/2})$.   

\smallskip\noindent
 {\bf Subregion A$\,$II:} $\beta>q$.

Now $\lld<\lmc^-$ and $1<y_-$; the next contribution to the
asymptotics beyond \(ldr12) comes from the square root singularity at
$y_-$, present in the second and third terms of \(chi2).  Now 
 \begin{equation}
 \Delta(y) = (1-y/y_-)(1-yp^2\lmc^-\lld), \label{ldIIdel}
 \end{equation}
 so that, writing $\rho_n^{(2)*}$ for the contribution to $\rho_n^{(2)}$
from this singularity, we have
 \begin{eqnarray}
 \rho_n^{(2)*}+\rho_n^{(3)} &=&
  \frac{1}{1+p\lld}
  \left[\ -\ \frac{\sqrt{1-yp^2\lmc^-\lld}}{2y(1-y)}
  \ +\  \frac{\beta(p-\beta)\sqrt{1-yp^2\lmc^-\lld}}{
    2yp^2(1-\beta)(\lhd-\lld y)}\right]_{y=y_-}\nonumber\\
   &&\hskip30pt \times \left(-\frac{y_-^{-n}}{2n\sqrt{\pi n}}\right)
   +O(y_-^{-n}/n^{5/2})
 \nonumber\\
 &=& -\ \frac{\alpha(p-\alpha)(1-p)^{1/4}\sqrt{\lmc^-}(\lld-\lhd)
    }{2(p-\alpha^2)(\lmc^- - \lld)(\lmc^- - \lhd)}\frac{1}{ n\sqrt{\pi n}}
   \left(\frac{\alpha(p-\alpha)}{(1-\alpha)(q+1-p)}\right)^n\nonumber\\
   &&\hskip30pt 
   +\ O(\left(\frac{1}{ n^2\sqrt{n}}
    \left(\frac{\alpha(p-\alpha)}{(1-\alpha)(q+1-p)}\right)^n\right).
   \label{ldr2*3}
 \end{eqnarray}
 From \(ldr12) and \(ldr2*3),
 \begin{eqnarray}
  \rho_n\ &=& \frac{\alpha(1-\alpha)}{ p-\alpha^2} \nonumber\\
   &&\hskip20pt
    - \ \frac{\alpha(p-\alpha)(1-p)^{1/4}\sqrt{\lmc^-}(\lld-\lhd)
    }{2(p-\alpha^2)(\lmc^- - \lld)(\lmc^- - \lhd)}\frac{1}{ n\sqrt{\pi n}}
   \left(\frac{\alpha(p-\alpha)}{(1-\alpha)(q+1-p)}\right)^n\nonumber\\
   &&\hskip20pt 
   +\ O\left(\frac{1}{ n^2\sqrt{n}}
    \left(\frac{\alpha(p-\alpha)}{(1-\alpha)(q+1-p)}\right)^n\right).
 \label{ldrII}
 \end{eqnarray}

 \smallskip\noindent
 {\bf The A$\,$I/A$\,$II boundary:} $\beta=q$.

Here $y_1=y_-$ and the leading correction to $\rho_n$ beyond the constant
term \(ldr12) is an $O(n^{-1/2})$ contribution from the third term in \(chi2).
From \(ldr12) and \(ldIIdel), 
 \begin{eqnarray}
   \rho_n &=&  \frac{\alpha(1-\alpha)}{ p-\alpha^2} 
  \ + \ \left[\frac{\beta(p-\beta)\sqrt{1-yp^2\lmc^-\lld}}{
   2(1+p\lld)yp^2(1-\beta)\lhd}\right]_{y=y_1}\frac{y_1^{-n}}{\sqrt{\pi n}}
    \ +\  O\left(\frac{y_1^{-n}}{ n\sqrt{\pi n}}\right)\label{ldrI-II}
  \\
  &=&  \frac{\alpha(1-\alpha)}{ p-\alpha^2} 
    \ + \ \frac{\alpha(p-\alpha)(1-p)^{1/4}}{
    (p-\alpha^2)}
    \sqrt\frac{1-\beta}{\beta(p-\beta)}\frac{1}{\sqrt{\pi n}}
   \left(\frac{\alpha(p-\alpha)}{ \beta^2(1-\alpha)}\right)^n
    \ +\  O\left(\frac{y_1^{-n}}{ n\sqrt{\pi n}}\right).\nonumber
 \end{eqnarray}

 The  low-density results \(ldrI), \(ldrII), and \(ldrI-II) agree with
\cite{DEHP} in the  $p\searrow0$ limit.

\smallskip\noindent
 (iii) {\bf The high density  region:} $\beta<\alpha$ and $\beta<q$
(region B in figure 2).

In this region the generating function is obtained by the substitution
$\lambda_0=\lhd$ in \(chi2).  After some tedious algebra, this leads to
 \begin{equation}
 \Phi(y) = \frac{p-\beta}{ p-\beta^2}\;\frac{1}{1-y},
 \end{equation}
 so that the density is constant:
 \begin{equation}
    \rho_n = \frac{p-\beta}{ p-\beta^2}.
 \label{hdd}
 \end{equation}
 %
\section{Exact Expressions for Finite Systems}
\label{combinatorial}
\setcounter{equation}{0}

In this section we obtain exact and explicit expressions for the
current and density profile for finite systems and all values of
$\alpha$, $\beta$ and $p$.  We shall do this by using the algebraic
rules (\ref{D1E1con}) and (\ref{D1V1}). This provides a complementary
approach to that of sections \ref{phasediagram} and \ref{asymptotics}
where large $N$ properties are calculated directly.

Our first task is  to evaluate
$ z_n=\langle W_1 | K^n | V_1 \rangle $.
We proceed by writing
$K^n$ as a sum of irreducible
(with respect to rule (\ref{D1E1con}))
strings in the following manner
\begin{equation}
K^n = \sum_{r=0}^{n} a_{n,r} \sum_{q=0}^{r} E_1^{r-q}D_1^{q}\quad.
\label{irred}
\end{equation}
It turns out that  $a_{n,r}$ is given by the expression
\begin{eqnarray}
\lefteqn{a_{n,r} =}&&\nonumber \\
&&\sum_{t=0}^{n-r} \left[
\left( \begin{array}{c} n\\r+t \end{array} \right)
\left( \begin{array}{c} n-r-1\\t \end{array} \right)
-
\left( \begin{array}{c} n+1\\r+t+1 \end{array} \right)
\left( \begin{array}{c} n-r-2\\t-1 \end{array} \right)
\right]
(1-p)^t\, .
\label{anr}
\end{eqnarray}
with the conventions 
$\left( \begin{array}{c} X\\0 \end{array} \right)=1$ 
 and
$\left( \begin{array}{c} X\\-1 \end{array} \right)=0$
for any integer $X$. The proof of (\ref{anr}) is left to 
appendix \ref{appendixb}.
Here we  check a few simple cases. 
From (\ref{anr}) and our conventions for the binomial
coefficients we find
\begin{eqnarray}
a_{n,n} &=& 1\quad, \nonumber \\
a_{n,n-1} &=& n-(1-p)\quad, \nonumber \\
a_{n,n-2} &=& 
\left( \begin{array}{c} n\\2 \end{array} \right) -(1-p)\quad,
\nonumber \\
a_{n,n-3} &=& 
\left( \begin{array}{c} n\\3 \end{array} \right) +
\left[ 2\left( \begin{array}{c} n\\2 \end{array} \right) 
-\left( \begin{array}{c} n+1\\2 \end{array} \right) \right](1-p)
-(1-p)^2\quad,
\end{eqnarray}
which yield using (\ref{irred})
\begin{eqnarray}
K &=& p + (D_1 +E_1)\quad,\nonumber\\
K^2&=& p + (1+p)(D_1 +E_1) + (D^2_1 +E_1 D_1 + E^2_1)\quad,\nonumber\\
K^3&=& 2p-p^2 + (2+p)(D_1 +E_1) + (2+p) (D^2_1 +E_1 D_1 + E^2_1)
\nonumber \\
&&
+(D_1^3 +E_1 D_1^2 +E_1^2 D_1+E_1^3)\quad,
\end{eqnarray}
as can be verified by direct calculation.
From  (\ref{anr}) we determine
an exact expression for $z_N$ by using the action
of $D_1,E_1$ (\ref{D1V1}):
\begin{equation}
z_N
= \sum_{r=0}^{N} a_{N,r} \frac{ 
(p(1-\beta)/\beta)^{r+1} - (p(1-\alpha)/\alpha)^{r+1} 
}{(p(1-\beta)/\beta) - (p(1-\alpha)/\alpha)}\quad,
\label{Kexp}
\end{equation}
where, without loss of generality, we have taken
$\langle W_1|V_1 \rangle=1$.
Together with (\ref{Jeq}), (\ref{Kexp}) yields an exact expression 
for the current.

To check the limit $p\to 0$
we use the identity
\begin{equation}
\sum_{i=-\infty}^\infty \left( \begin{array}{c} N \\ X-i \end{array} \right)
\left( \begin{array}{c} M \\ Y+i \end{array} \right)
= \left( \begin{array}{c} N+M \\ X+Y \end{array} \right)
\label{ident3}
\end{equation}
in order  to obtain
\begin{eqnarray}
a_{n,r}&\to&
\left( \begin{array}{c} 2n-r-1\\n-r \end{array} \right)
-
\left( \begin{array}{c} 2n-r-1\\n-r-1 \end{array} \right) 
\nonumber \\
&=&
 \frac{r (2n-r-1)!}{n! (n-r)!}\; .
\end{eqnarray}
This  agrees with Eq. 39 of \cite{DEHP}.

Also consider $p=1$,
then (\ref{anr}) becomes
$a_{n,r}= 
\left( \begin{array}{c} n\\r \end{array} \right)
$ and
\begin{equation}
z_N
= \frac{ (1-\beta)\beta^{-(N+1)}
-(1-\alpha)\alpha^{-(N+1)}
}{(1-\beta)/\beta - (1-\alpha)/\alpha}\quad.
\end{equation}
One can check that this recovers the results (\ref{zp=1a}) and
(\ref{zp=1b})  of section \ref{pequalone}.

In order to write down the density profile we  use an expression derived
in appendix \ref{appendixb}:
\begin{equation}
D_1K^n =(1-p) \sum_{r=0}^{n-1} A(r)K^{n-r}
+ \sum_{r=0}^{n} a_{n,r} D_1^{r+1}\,,
\label{DKn}
\end{equation}
where $a_{n,r}$ is given by (\ref{anr}) and
\begin{equation}
A(m) = \sum_{t=0}^{m-1} \frac{1}{m}
\left( \begin{array}{c} m\\t \end{array} \right)
\left( \begin{array}{c} m\\t+1 \end{array} \right)
 (1-p)^t\, ,
\label{Adef}
\end{equation}
with the convention  $A(0)=1$.
It can be checked that
\begin{equation}
p A(n-1) = a_{n,0}\;\;\;\mbox{for}\;\;\;n>0\quad.
\label{pAa}
\end{equation}
Inserting (\ref{DKn}) into (\ref{density}) yields
\begin{equation}
\langle \tau_i \rangle_N = Z_N^{-1}\times 
\left[ pz_{N-1}+(1-p) \sum_{r=0}^{N-i-1} A(r) z_{N-1-r}+
z_{i-1}  \sum_{r=0}^{N-i} a_{N-i,r} \left( p(1-\beta)/\beta \right)^{r+1} \right]\,. 
\label{tauexact}
\end{equation}

Expression (\ref{tauexact}) together with (\ref{Kexp}) gives an exact
expression for the density profile of parallel updating
for all system sizes.  
Through the mappings of section \ref{mappings} it also
provides   exact expressions for the density profiles
of ordered and sublattice parallel updating.

\bigskip
\noindent {\bf The Case $\alpha = \beta =p$}

More can be said in the special case of $\alpha = \beta =p$
where many formulae simplify considerably. We take advantage of this
to  simplify the expression for the density profile and to write
the two-point correlation functions as a sum of one point correlation
functions.

First we note that in this case (\ref{Kexp})  simplifies as follows:
\begin{eqnarray}
z_N &=& \sum_{r=0}^{n}
\sum_{t=r}^{n} \left[
\left( \begin{array}{c} n\\t \end{array} \right)
\left( \begin{array}{c} n-r-1\\t-r \end{array} \right)
-
\left( \begin{array}{c} n+1\\t+1 \end{array} \right)
\left( \begin{array}{c} n-r-2\\t-r-1 \end{array} \right)
\right] (r+1) (1-p)^t 
\nonumber \\
&=&\sum_{t=0}^{n} \left[
\left( \begin{array}{c} n\\t \end{array} \right)
\left( \begin{array}{c} n+1\\t \end{array} \right)
-
\left( \begin{array}{c} n+1\\t+1 \end{array} \right)
\left( \begin{array}{c} n\\t-1 \end{array} \right)
\right] (1-p)^t 
\nonumber \\
&=&
\sum_{t=0}^{n} \frac{1}{n+1}
\left( \begin{array}{c} n+1\\t \end{array} \right)
\left( \begin{array}{c} n+1\\t+1 \end{array} \right)
 (1-p)^t=A(n+1)\, ,
\label{Kp}
\end{eqnarray}
where $A(m)$ is given by (\ref{Adef}) and we have used
\begin{equation}
\sum_{r=0}^t
\left( \begin{array}{c} M-r\\t-r \end{array} \right)(r+1)
=\left( \begin{array}{c} M+2\\t \end{array} \right)\, .
\end{equation}
 From our mapping of the model onto the ASEP with a second class
particle, described in section \ref{mappings}, one can check
that (\ref{Kp}) is precisely formula (4.10) in \cite{DJLS}. 

We also find using the representation
(\ref{W1V1})--(\ref{E1}) that
\begin{equation}
(D_1 +p) K -K (D_1+p)= K(E_1+p)-(E_1+p)K= D_1E_1-E_1D_1 =
 (1-p) |V_1 \rangle  \langle W_1|\, .
\label{projector}
\end{equation}
Thus the density profile is given by
\begin{eqnarray}
\langle \tau_i \rangle_N
&=& \langle \tau_{i+1} \rangle_N + \frac{(1-p)}{Z_N}
A(i)A(N-i-1) \nonumber \\
&=&  \frac{(1-p) \sum_{n=0}^{N-i} A(N-n)A(n) +p A(N)
        }{A(N+1)+pA(N)}\, .
\end{eqnarray}

One can also use (\ref{projector}) to relate
the two-point correlation function (\ref{twopoint})
to one-point  correlations:
\begin{eqnarray}
\langle \tau_i (1- \tau_j) \rangle_N
&=&
\langle \tau_i (1- \tau_{j-1}) \rangle_N
+ \frac{(-p)^{j-i-1}}{Z_N} A(N-j+i+1)
\nonumber \\
&&
+(1-p)\frac{A(N-j+1)}{Z_N}
\sum_{n=0}^{j-i-2}(-p)^n
\langle \tau_i \rangle_{j-2-n} Z_{j-2-n} \\
&=&
\frac{1}{Z_N}\sum_{l=i+1}^{j}(-p)^{l-i-1}  A(N-l+i+1)
\nonumber \\
&&+  \frac{(1-p)}{Z_N}
\sum_{l=i+2}^{j} A(N-l+1) \sum_{n=0}^{l-i-2}(-p)^n
\langle \tau_i \rangle_{l-2-n} Z_{l-2-n}\quad.
\end{eqnarray}
\section{The density in the bulk}
\label{bulk}
\setcounter{equation}{0}

In this section we combine the information derived from generating functions
in sections~\ref{phasediagram} and \ref{asymptotics} with the the exact
calculations of the preceding section to obtain expressions for the bulk
density of the system, that is, for the large-$N$ limit of $\langle \tau_i
\rangle_N$ at constant $\theta = i/N$.  As we will see, this bulk density is
constant except on the boundary of the low and high density regions, and its
value is may be guessed by taking the $n\to \infty$ limit in the asymptotics
of section~\ref{asymptotics}, that is, in (\ref{asymc}), in (\ref{ldrI}) and
(\ref{ldrII}), and in (\ref{hdd}) (where no limit is needed).  However, it
needs to be shown that this limit indeed gives the correct bulk density, and
we shall do so here. 

The key to the calculation is to study the difference in densities at
successive sites.  Writing $z_n=z_n(\alpha,\beta,p)$,
$z_n^*=z_n(1,\beta,p)$, and $z_n^{\dagger}=z_n(p,p,p)$, we have 
from (\ref{tauexact}), (\ref{Kexp}), and (\ref{Kp}),
 \begin{equation}
 \langle \tau_i \rangle_N = \frac
 { pz_{N-1}(\alpha,\beta,p) 
   + (1-p)\sum_{r=0}^{N-i-1} z^{\dagger}_{r-1}z_{N-1-r}
  + [p(1-\beta)/\beta]z_{i-1}z^*_{N-i}}
   {z_N+pz_{N-1}}\,,
 \label{tauexact2}
 \end{equation}
 and hence the density difference
 $ \Delta\rho_i=\langle \tau_i \rangle_N -\langle \tau_{i-1} \rangle_N $ is
given by 
 \begin{equation}
 \Delta\rho_i
  = \frac {[p(1-\beta)/\beta](z_{i-1}z^*_{N-i} - z_{i-2}z^*_{N-i+1})
   - (1-p)z^{\dagger}_{N-i-1}z_{i-1} }
   {z_N+pz_{N-1}}\,.
 \label{taudiff}
 \end{equation}
 We analyze the asymptotic ($N\to\infty$) behavior of (\ref{taudiff}) in
various parts of the phase plane; properties of the bulk are obtained by the
scaling $i=\theta N$, $0<\theta<1$, and we will also be interested in the
transition regions $i\ll N$, $N-i\ll N$.  The boundary regions in which $i$
or $N-i$ remains finite were investigated in section~\ref{asymptotics}. 

The point $\alpha=\beta=p$ always lies in the maximum current region, so that
by (\ref{znmc}), $z^\dagger_n\sim Cn^{-3/2}(\lmc^-)^{-n}$ for large $n$,
where here and below $C$ designates some unspecified constant.  When
$(\alpha,\beta,p)$ lies in the maximum current region, so does $(1,\beta,p)$,
so that $z_n$ and $z^*_n$ have this same asymptotic form and thus for $N$,
$i$, and $N-i$ large, $\Delta\rho_i\sim CN^{3/2}i^{-3/2}(N-i)^{-3/2}$.  Thus
in the bulk $\Delta\rho_i=O(N^{-3/2})$ and hence the density
in the bulk is constant.  Moreover, 
 \begin{equation}
  \langle \tau_{N-i} \rangle_N -\rho_{\rm bulk}
 = \sum_{j=\theta N}^{N-i}\Delta\rho_j
 \end{equation}
 vanishes as $i,N\to\infty$, so that (\ref{rhomc}) implies that this bulk
density has value $1/2$.  

When $(\alpha,\beta,\gamma)$ lies in the low density region and $\beta<q$
(i.e., in subregion I), $(1,\beta,p)$ lies in the high density region, and
thus from (\ref{znld}) and reflection symmetry, $z_n\sim C\lld^{-n}$ and
$z^*_n\sim C\lhd^{-n}$; since here $\lhd<\lmc^-$, $\Delta\rho_i\sim
C(\lld/\lhd)^{(N-i)}$.  As above, this implies that the bulk density is constant
and equal to its value at the left end of the system, which, from
(\ref{hdd}) and the reflection symmetry, is $\alpha(1-\alpha)/(p-\alpha^2)$. 
The argument in subregion II is similar, with $z^*_n\sim
Cn^{-3/2}(\lmc^-)^{-n}$ and
 $\Delta\rho_i\sim C(N-i)^{-3/2}(\lld/\lmc^-)^{(N-i)}$; the bulk density is
the same.  By reflection symmetry the bulk density in the high
density phase is constant and equal to $(p-\beta)/(p-\beta^2)$.

The case $\alpha=\beta<q$ requires special attention. Here a slight extension
of the arguments of section~\ref{phasediagram} shows that
 $z_n\sim Cn\lld^{-n}$.  As in the subregion I case above,
 $z^*_n\sim C\lhd^{-n}$ and the $z_n^\dagger$ term in
(\ref{taudiff}) can be neglected; since $\lhd=\lld$, 
 $\Delta\rho_i\sim C N^{-1}$, so that the density profile is linear.  The
values of $\langle\tau_i\rangle_N$ at the left and right ends of the system
are, from (\ref{hdd}) and the symmetry, $\alpha(1-\alpha)/(p-\alpha^2)$ and
$(p-\beta)/(p-\beta^2)$, respectively.  This linear profile may, as usual, be
interpreted as a superposition of shocks.
\section{Summary}
\label{summary}

In this paper, we have presented an exact solution for the steady state of a
simple cellular automaton describing traffic flow: the ASEP with parallel
(synchronous) updating and open boundary conditions.  The solution is based
on recursion relations in the system size for the steady-state weights of the
configurations, or, equivalently, on formulae for these weights as matrix
elements of operators satisfying a quartic algebra.  By writing these
operators as rank four tensors, we were also able to express the relevant
physical quantities in terms of a simpler matrix algebra in which the 
operators satisfy quadratic relations. 

We used several different methods to extract explicit expressions for
observables from the matrix algebra.  The first applied when $p=1$, in
which case a two dimensional representation of the quadratic algebra
exists; this made it possible to obtain analytic expressions, in both
the finite and the infinite system, for the current and for one and
two point correlation functions.  The results confirmed conjectures in
\cite{RSSS,TE}.  The two point function is particularly interesting,
because its oscillating behavior directly reflects the particle-hole
attraction caused by the parallel updating.  For $\alpha=\beta$ (still
with $p=1$) we obtained also closed formulae for the fluctuations in
the number of particles (cars) in the system.  Second, for general $p$
we derived exact formulae, in finite systems, for the current and the
one point function, and for the two point function in the case
$\alpha=\beta=p$; the method was essentially an inductive use of the
relations of the matrix algebra.  The resulting formulae involve
rather complicated combinatorial expressions in which it is difficult
to take the limit of infinite system size.  Third, again for general
$p$, we used the analytic properties of generating functions to
compute asymptotic expressions for the current (and therefore the
phase diagram) and for density profiles near the boundaries of the
system.  Finally, we combined the results of the last two methods to
determine the density in the bulk.

Our results confirm the phase diagram conjectured in \cite{RSSS}.  It is
similar to that of random sequential updating \cite{DEHP,SD}: there are
three phases and, for example, near the right boundary we found exponential
decay to the bulk density in the low density phase, algebraic decay to the
bulk density in the maximum current phase, and a constant density profile
in the high density phase.  As $p$ increases, the portion of the phase
plane corresponding to the maximum current phase shrinks until at $p=1$
only the high and low density phases are present. It would be of interest
to see if the phase diagram could be predicted by simple physical
considerations such as those of \cite{KSKS}.

By considering mappings of the matrix algebra used here to those
applicable to other updating schemes, we can directly translate all
our results (finite size and asymptotics) for the current and the
profiles to the case where the update of the ASEP is done in discrete
time but not simultaneously (specifically, with the ordered sequential
update and sublattice parallel updates).  A similar mapping of
algebras shows that our results apply to a system of particles on a
ring, with one second class particle, in the grand canonical ensemble.
Since relatively few exact properties of the discrete time updating
schemes were previously known---essentially only the asymptotic
current \cite{honi}---we obtain new results for these models. For
example, we verify all conjectured results in Table I of \cite{RSSS}
(these describe bulk properties) and derive new formulae for density
profiles both for finite systems and asymptotically. The simple
translation rules for the current and one-point functions
(independent of the system size or any other parameters) are
surprising, since the two point function of the ASEP with
parallel updating is very different from that for the other updating
schemes.  It would be interesting to investigate if similar relations
are true not only for the ASEP but for other models.
\quad
 \par
 \vskip20pt
\noindent
{\Large\bf Acknowledgments}
\quad
 \par
\vskip10pt
 \noindent 
MRE is a Royal Society University Research Fellow and thanks DIMACS and its
supporting agencies, the NSF under contract STC--91--19999 and the NJ
Commission on Science and Technology, for support.  NR gratefully
acknowledges a postdoctoral fellowship from the Deutsche
Forschungsgemeinschaft and thanks Joel Lebowitz for hospitality at the
Mathematics Department of Rutgers University and for support under NSF
grant DMR~95--23266.  We thank Ovidiu Costin for useful discussions.

\appendix\def\theequation{A.\arabic{equation}}
\section {Asymptotics from generating functions}
\label{appendixa}
\setcounter{equation}{0}

The asymptotic behavior of the coefficients $h_n$ of a generating function
$h(t)=\sum_{n=0}^\infty h_n t^n$ can frequently be determined, up to order
$o(s^{-n})$, from knowledge of the singularities of $h$ in a disk $|t|<r$
with $r>s$.  We analyze below the two cases of this sort.  Recall that for
any real $\gamma$ and complex number $t^*$ we have the following Taylor series,
 \begin{equation}
   (1-t/t^*)^\gamma = \sum_{n=0}^{\infty} a_{\gamma,n}
\left(\frac{t}{ t^*}\right)^n,\qquad
       a_{\gamma,n} = O(n^{-\gamma-1})\;.\label{gamma}
 \end{equation}
For application to sections \ref{phasediagram}
and \ref{asymptotics} we will  need the special cases
 \begin{eqnarray}
   a_{-1,n} &=& 1\;,\label{mone}\\
   a_{-1/2,n} &=& \frac{1}{\sqrt{\pi n}}-\frac{1}{8n\sqrt{\pi n}}
         +O\left(\frac{1}{ n^2\sqrt{n}}\right)\;, \label{mhalf}\\
   a_{1/2,n} &=& -\frac{1}{ 2n\sqrt{\pi n}}
     +O\left(\frac{1}{ n^2\sqrt{n}}\right)\;.\label{half}
 \end{eqnarray}

 \smallskip\noindent
 {\bf Case 1:} If the only singularities of $h$ in the disc
$|z|<r$ are simple poles at $t_1,\ldots,t_m$ and
$c_i=\lim_{t\to t_i}(1-t/t_i)h(t)$, then
$h-\sum_{i=1}^m c_i(1-t/t_i)^{-1}$ is analytic in $|t|<r$ and hence
for any $s<r$ we have, from \(gamma) and \(mone),
 \begin{equation}
  h_n= c_1t_1^{-n}+\cdots + c_mt_m^{-n} + o(s^{-n}). \label{simple}
 \end{equation}

 \smallskip\noindent
 {\bf Case 2:} Suppose that $h(t)$ has simple poles at $t_i$, with $c_i$
defined as above, as well as a power singularity at some point $t_0>0$; we
assume that $|t_i|<t_0$ for $i=1,\ldots,m$ and that $h(t)$ can be written in
the form 
\begin{equation}
h(t)=g(t)(1-t/t_0)^\gamma\;,
\end{equation}
 where the only singularities of $g$ in
the disk are the poles.  Then for any $k\ge0$,
 \begin{equation}
   h_n= \sum_{i=0}^mc_it_i^{-n}
     + \sum_{j=0}^k b_j a_{\gamma+j,n}t_0^{-n}
   + O(n^{-\gamma-k-2}t_0^{-n}) \label{alg}
 \end{equation}
 where $b_j=g^{(j)}(t_0)(-t_0)^j/j!$.  To verify \(alg) write $h_n=(2\pi
i)^{-1}\oint_Ch(t)t^{-n-1}\,dt$, where the contour $C$ has $m+1$ components
$C_0,C_1,\ldots C_m$.  For $i\ge1$, $C_i$ is a small circle, traced
clockwise, around the point $t_i$, and gives the term $c_it_i^{-n}$ in
\(alg).  $C_0$ follows the circle $|t|=s$ counter-clockwise from just
above to just below the positive real axis, then the real axis to
$t_0+\epsilon$ (for $\epsilon$ very small), then a small circle of radius
$\epsilon$ clockwise around $t=t_0$, then the real axis to $t=s$.  In
evaluating the integral on $C_0$ we choose $k$ so that $k+\gamma>0$ (proving
the result for such a $k$ proves it also for smaller $k$) and write
$g(t)=\sum_{j=0}^k b_j(1-t/t_0)^j+g_{k+1}(t)(1-t/t_0)^{k+1}$.  The
contribution of $\oint_{C_0} b_j(1-t/t_0)^{j+\gamma}t^{-n-1}\,dt$ is
precisely $b_ja_{\gamma+j,n}t_0^{-n}$. There are two terms in 
the remaining contribution:  the 
integral over the large circle is $O(s^{-n})$, while the integral back and
forth over the
real axis and around the small circle is, by our choice of $k$, a constant
multiple of $\int_{t_0}^s g_{k+1}(t)(1-t/t_0)^{k+1+\gamma}t^{-n-1}\,dt$,
which is easily estimated by the saddle point
method to be $O(n^{-\gamma-k-2}t_0^{-n}) $. 
\def\theequation{B.\arabic{equation}}
\section {Proof of formulae (\ref{anr}) and (\ref{DKn})}
\label{appendixb}
\setcounter{equation}{0}

{\it Proof of (\ref{anr})}: \\
In this appendix we shall prove
\begin{equation}
K^n = \sum_{r=0}^{n} a_{n,r} \sum_{q=0}^{r} E_1^{r-q}D_1^{q}
\label{irredapp}
\end{equation}
where  $a_{n,r}$ is given by the following
expression
\begin{eqnarray}
\lefteqn{a_{n,r} =}&&\nonumber \\
&&\sum_{t=0}^{n-r} \left[
\left( \begin{array}{c} n\\n-r-t \end{array} \right)
\left( \begin{array}{c} n-r-1\\t \end{array} \right)
-
\left( \begin{array}{c} n+1\\n-r-t \end{array} \right)
\left( \begin{array}{c} n-r-2\\t-1 \end{array} \right)
\right]
(1-p)^t\;.
\label{anrapp}
\end{eqnarray}

We  first require a preliminary result:
\begin{equation}
(D_1+p)E_1^n = (1-p)^n( D_1 +p) 
+ \sum_{m=0}^{n-1}(1-p)^m E_1^{n-m}
\label{DEn}
\end{equation}
which is easy to prove by induction using
(\ref{D1E1con}): for $n=1$ one has
$(D_1+p)E_1 = (1-p)( D_1 +p) +E_1$;
then assuming (\ref{DEn}) and right multiplying by $E_1$ yields
\begin{eqnarray}
(D_1 +p) E^{n+1} &=&
(1-p)^n\left[ (1-p)( D_1 +p) +E \right]
+ \sum_{m=2}^{n+1}(1-p)^{m} E_1^{n-m+1}\\
 &=&
(1-p)^{n+1}\left[  D_1 +p \right]
+ \sum_{m=1}^{n+1}(1-p)^{m} E_1^{n-m+1}\;,
\end{eqnarray}
hence (\ref{DEn}) is proven by induction.

Using (\ref{DEn}) 
eventually leads to the following recursion for
$a_{n,r}$
\begin{eqnarray}
a_{n+1,r}&=&a_{n, r-1}
+\sum_{m=0}^{n-r}a_{n, r+m}(1-p)^m\;\;\;\mbox{for}\;\;\; 1 \leq r \leq n
\label{arecur1}\\
a_{n+1,0}&=&
\sum_{m=0}^{n}a_{n, m}p(1-p)^m
\label{arecur2}\\
a_{n+1,n+1}&=&a_{n,n}
\label{arecur3}
\end{eqnarray}
with boundary condition $a_{0,0}=1$.
To see this left multiply (\ref{irredapp}) by $K$
\begin{eqnarray}
K^{n+1}&=& (E_1 +D_1+p)
\sum_{r=0}^{n} a_{n,r} \sum_{q=0}^{r} E_1^{r-q}D_1^{q} 
 \\
&=&\sum_{r=1}^{n+1} a_{n,r-1} \sum_{q=0}^{r-1} E_1^{r-q}D_1^{q}
+\sum_{r=0}^{n} a_{n,r} \sum_{q=0}^r(1-p)^{r-q} \left[ D_1 +p \right]D_1^{q}
\nonumber  \\
&&+\sum_{r=1}^{n} a_{n,r} \sum_{q=0}^{r-1}
 \sum_{m=0}^{r-q-1}(1-p)^{m}E_1^{r-q-m} D_1^{q}
\label{Kn1}
\end{eqnarray}
where we have relabeled the indices $r,q$ in the first term of
(\ref{Kn1}) and used (\ref{DEn}) to generate the second two
terms.
Relying  on not a little  dexterity in relabeling and manipulating
sums one can develop the second two terms of
(\ref{Kn1}) as follows
\begin{eqnarray}
\lefteqn{\sum_{r=0}^{n} a_{n,r} \sum_{q=0}^r(1-p)^{r-q}
 \left[ D_1 +p \right]D_1^{q}}
\nonumber && \\
&&=
\sum_{r=0}^{n} a_{n,r} \left[(1-p)^r p + D_1^{r+1} \right]
+\sum_{r=1}^{n} a_{n,r} \sum_{q=1}^{r}
 \left[ (1-p)^{r+1-q} D_1^q +(1-p)^{r-q} pD_1^q  \right]
\nonumber \\
&&=
\sum_{r=0}^{n} a_{n,r} \left[(1-p)^r p + D_1^{r+1} \right]
+\sum_{r=1}^{n} a_{n,r}\sum_{q=1}^{r}
(1-p)^{r-q} D_1^q
\nonumber \\
&&=
\sum_{r=0}^{n} a_{n,r} \left[(1-p)^r p + D_1^{r+1} \right]
+\sum_{q=1}^{n} \sum_{r=0}^{n-q}a_{n,r+q}
(1-p)^{r} D_1^q
\label{Kn2}
\end{eqnarray}
and
\begin{eqnarray}
\lefteqn{ \sum_{r=1}^{n} a_{n,r} \sum_{q=0}^{r-1}
 \sum_{m=0}^{r-q-1} (1-p)^{m} E_1^{r-q-m} D_1^{q} }
\nonumber && \\
&&=\sum_{m=0}^{n-1}  \sum_{r=m+1}^{n} \sum_{q=0}^{r-m-1}a_{n,r}
(1-p)^{m}E_1^{r-q-m} D_1^{q}
\nonumber \\
&&=\sum_{m=0}^{n-1}  \sum_{r=1}^{n-m} \sum_{q=0}^{r-1} a_{n,r+m}
(1-p)^{m}E_1^{r-q} D_1^{q}
\nonumber \\
&&= \sum_{r=1}^{n} \sum_{q=0}^{r-1} \sum_{m=0}^{n-r} a_{n,r+m}
(1-p)^{m} E_1^{r-q} D_1^{q}\;.
\label{Kn3}
\end{eqnarray}
When the expressions (\ref{Kn3}), (\ref{Kn2}) are inserted back into
(\ref{Kn1}), the second term in the square brackets of (\ref{Kn2})
becomes the $q=r$ component of the first term of (\ref{Kn1}), and
after relabeling indices the third term of (\ref{Kn2}) becomes the
$q=r$ component of (\ref{Kn3}), leading to
\begin{eqnarray}
K^{n+1}&=&
\sum_{r=1}^{n+1} \sum_{q=0}^{r} a_{n,r-1}  E_1^{r-q} D_1^{q} 
\label{Kn4} \\
&&
+\sum_{r=1}^{n} \sum_{q=0}^{r-1} \sum_{m=0}^{n-r}  a_{n,r+m}
(1-p)^m E_1^{r-q} D_1^{q} 
+\sum_{m=0}^{n}  p(1-p)^m a_{n,m} \;.
\nonumber
\end{eqnarray}
From (\ref{Kn4}) one can read off
(\ref{arecur1})--(\ref{arecur3}).

Now assume that  $a_{n,r}$ can be written as
\begin{equation}
a_{n,r} = \sum_{t=0}^{n-r} d_{n,r,t}(1-p)^t\;.
\label{dnrt}
\end{equation}
Inserting (\ref{dnrt}) into
(\ref{arecur1}), (\ref{arecur2}) and (\ref{arecur3})
respectively yields
\begin{eqnarray}
d_{n+1,r,t}&=&d_{n,r-1,t}+\sum_{m=0}^t d_{n,r+m,t-m}
\;\;\;\mbox{for}\;\;\;1\leq r \leq N
\label{drecur1}\\
d_{n+1,0,t}&=& \sum_{m=0}^{t} d_{n,m,t-m}
-\sum_{m=0}^{t-1} d_{n,m,t-1-m} \label{drecur2} \\
d_{n,n,t}&=& \delta_{t,0} \label{drecur3}
\end{eqnarray}
In order to show that 
\begin{equation}
d_{n,r,t}=
\left( \begin{array}{c} n\\r+t \end{array} \right)
\left( \begin{array}{c} n-r-1\\t \end{array} \right)
-
\left( \begin{array}{c} n+1\\r+t+1 \end{array} \right)
\left( \begin{array}{c} n-r-2\\t-1 \end{array} \right)
\label{dexp}
\end{equation}
satisfies (\ref{drecur1})--(\ref{drecur3})
we employ two well known identities
\begin{equation}
\sum_{i=0}^{N-M} \left( \begin{array}{c} N-i \\ M -i \end{array} \right)
= \left( \begin{array}{c} N+1 \\ M \end{array} \right)
\label{ident1}
\end{equation}
\begin{equation}
\left( \begin{array}{c} N \\ M \end{array} \right)=
\left( \begin{array}{c} N-1 \\ M \end{array} \right)
+
\left( \begin{array}{c} N-1 \\ M-1 \end{array} \right)
\label{ident2}
\end{equation}
Using (\ref{ident1}) yields
\begin{eqnarray}
\sum_{m=0}^t d_{n,r+m,t-m} &=&
\left( \begin{array}{c} n\\r+t \end{array} \right)
\left( \begin{array}{c} n-r\\t \end{array} \right)
-
\left( \begin{array}{c} n+1\\r+t+1 \end{array} \right)
\left( \begin{array}{c} n-r-1\\t-1 \end{array} \right)
\label{dsum}
\end{eqnarray}
Then (\ref{ident2}) becomes
\begin{eqnarray}
\lefteqn{ d_{n,r-1,t}+\sum_{m=0}^t d_{n,r+m,t-m} =}\nonumber &&\\
&&
\left( \begin{array}{c} n+1\\r+t \end{array} \right)
\left( \begin{array}{c} n-r\\t \end{array} \right)
-
\left( \begin{array}{c} n+2\\r+t+1 \end{array} \right)
\left( \begin{array}{c} n-r-1\\t-1 \end{array} \right)
\end{eqnarray}
which is the expression (\ref{dexp}) required to satisfy
(\ref{drecur1}).
Similarly with the aid of (\ref{ident1}) then repeated use of
(\ref{ident2}) one finds
\begin{eqnarray}
\lefteqn{ \sum_{m=0}^t d_{n,m,t-m} 
- \sum_{m=0}^{t-1} d_{n,m,t-1-m} }\nonumber &&\\
=&&
\left( \begin{array}{c} n\\t \end{array} \right)
\left( \begin{array}{c} n\\t \end{array} \right)
-
\left( \begin{array}{c} n+1\\t+1 \end{array} \right)
\left( \begin{array}{c} n-1\\t-1 \end{array} \right)
-
\left( \begin{array}{c} n\\t-1 \end{array} \right)
\left( \begin{array}{c} n\\t-1 \end{array} \right)
+
\left( \begin{array}{c} n+1\\t \end{array} \right)
\left( \begin{array}{c} n-1\\t-2 \end{array} \right)
\nonumber  \\
=&&
\left( \begin{array}{c} n+1\\t \end{array} \right)
\left( \begin{array}{c} n\\t \end{array} \right)
-
\left( \begin{array}{c} n+2\\t+1 \end{array} \right)
\left( \begin{array}{c} n-1\\t-1 \end{array} \right)
\end{eqnarray}
thus satisfying (\ref{drecur2}) when $d_{n,r,t}$ is given by
(\ref{dexp}).
Finally
when  the conventions
$\left( \begin{array}{c} X\\0 \end{array} \right)=1$ 
 and
$\left( \begin{array}{c} X\\-1 \end{array} \right)=0$
$\forall X$ are imposed,
 (\ref{dexp}) implies
\begin{equation}
d_{n,n,t}= \left( \begin{array}{c} n+1\\n+1+t \end{array} \right)
\left( \begin{array}{c} -1\\t \end{array} \right)
-
\left( \begin{array}{c} n+1\\-t-1 \end{array} \right)
\left( \begin{array}{c} -2\\t-1 \end{array} \right)
=\delta_{t,0}
\end{equation}
thus satisfying (\ref{drecur3}).

\quad\\ {\it Proof of (\ref{DKn})} \\
Here we prove
\begin{equation}
D_1K^n =(1-p) \sum_{r=0}^{n-1} A(r)K^{n-r}
+ \sum_{r=0}^{n} a_{n,r} D_1^{r+1}\,,
\label{DKnapp}
\end{equation}
First we note  
\begin{equation}
D_1^n \left[ E_1 +p \right] =
 (1-p)^n \left[ E_1 +p  \right] 
+ \sum_{m=0}^{n-1} (1-p)^{m} D_1^{n-m}
\label{DnK}
\end{equation} 
which is proven in a similar fashion to (\ref{DEn}).

To prove (\ref{DKnapp}) by induction, one can check the case $n=0$ or
$n=1$, then right multiply the rhs of (\ref{DKnapp}) by $K$,
 using (\ref{DnK}) to obtain
\begin{eqnarray}
D_1 K^{n+1} &=& (1-p) \sum_{r=0}^{n-1} A(r) K^{n+1-r}
+\sum_{r=0}^{n} a_{n,r} D_1^{r+1} \left[ D_1 +E_1 +p \right]
\nonumber \\
&=& (1-p)\sum_{r=0}^{n-1} A(r) K^{n+1-r}
+ \sum_{r=0}^n a_{n,r} D_1^{r+2} + \sum_{r=0}^n a_{n,r}(1-p)^{r+1} (E_1+p)
\nonumber \\
&&+ \sum_{r=0}^n \sum_{m=0}^{r} a_{n,r} (1-p)^{m} D_1^{r+1-m} 
\label{DKn1}
\end{eqnarray}
The third term of (\ref{DKn1}) becomes using (\ref{arecur2})
\begin{equation}
\sum_{r=0}^n a_{n,r}(1-p)^{r+1} (E_1+p) =
\frac{1-p}{p} a_{n+1,0} (E_1 +p)\;.
\label{DKn2}
\end{equation}
The fourth term of (\ref{DKn1}) may be developed as follows
\begin{eqnarray}
\sum_{r=0}^n \sum_{m=0}^{r} a_{n,r} (1-p)^{m} D_1^{r+1-m} 
=
\sum_{m=0}^n \sum_{r=0}^{n-m} a_{n,r+m} (1-p)^{m} D_1^{r+1}
\nonumber \\
=
\sum_{r=0}^n \sum_{m=0}^{n-r} a_{n,r+m} (1-p)^{m} D_1^{r+1}
=
\frac{a_{n+1,0}}{p} D_1 + \sum_{r=1}^{n}
\left[ a_{n+1,r}- a_{n,r-1}\right] D_1^{r+1}
\label{DKn3}
\end{eqnarray}
where (\ref{arecur1}), (\ref{arecur2}) have been used to obtain the final
equality.
Putting (\ref{DKn1}), (\ref{DKn2}) and (\ref{DKn3}) together yields
\begin{eqnarray}
D_1 K^{n+1} &=& (1-p) \sum_{r=0}^{n-1} A(r) K^{n+1-r}
+\frac{1-p}{p} a_{n+1,0}K + a_{n+1,0}D_1
+\sum_{r=1}^{n} a_{n+1,r} D_1^{r+1}
+D_1^{n+2}\nonumber\\ 
&=& (1-p) \sum_{r=0}^{n} A(r) K^{n+1-r}
+ \sum_{r=0}^{n+1} a_{n+1,r} D_1^{r+1}
\end{eqnarray}
which agrees with (\ref{DKnapp}), thereby
 proving (\ref{DKn})
by induction.


\newpage
\setcounter{equation}{0}
\newcommand{\graphik}[4]{\begin{figure}[t]\vspace{0.5cm}\begin{center}
\leavevmode\epsfig{file=#1, height=#2}\caption{{\small\it #3 \label{#4}}}
\end{center}\end{figure}}
\begin{center}
{\large Figure Captions}
\end{center}
\begin{description}
\item[Fig.\ 1] The exact truncated correlation function 
$g(i,j)$ in the case $p=1$,
for $i=25$ versus $j$. The system size is $N=100$. The oscillating
curve (squares) was obtained for $\alpha=\beta=0.9$, the other curve
(crosses) for $\alpha=\beta=0.1$.
\item[Fig.\ 2] (taken from \cite{RSSS}): Phase diagram for the ASEP with 
parallel (synchronous) update 
for $p=0.5$. C is the maximum current
phase, A and B are the low and high density phase, respectively.
The straight dashed lines are the boundaries between 
phase A$\,$I and A$\,$II (B$\,$I and B$\,$II). The curved dashed line
is the line given by (\ref{specialline}) and intersects the line
$\alpha=\beta$ at $\alpha=\beta=1-\sqrt{1-p}=q$ (see section
\ref{phasediagram}).
The inserts show typical 
density profiles in the various phases; note that the profile is
qualitatively the same in region A$\,$I (B$\,$I) 
and in the portion of region A$\,$II (B$\,$II)
below the curved dashed line.
\end{description}
\newpage
\graphik{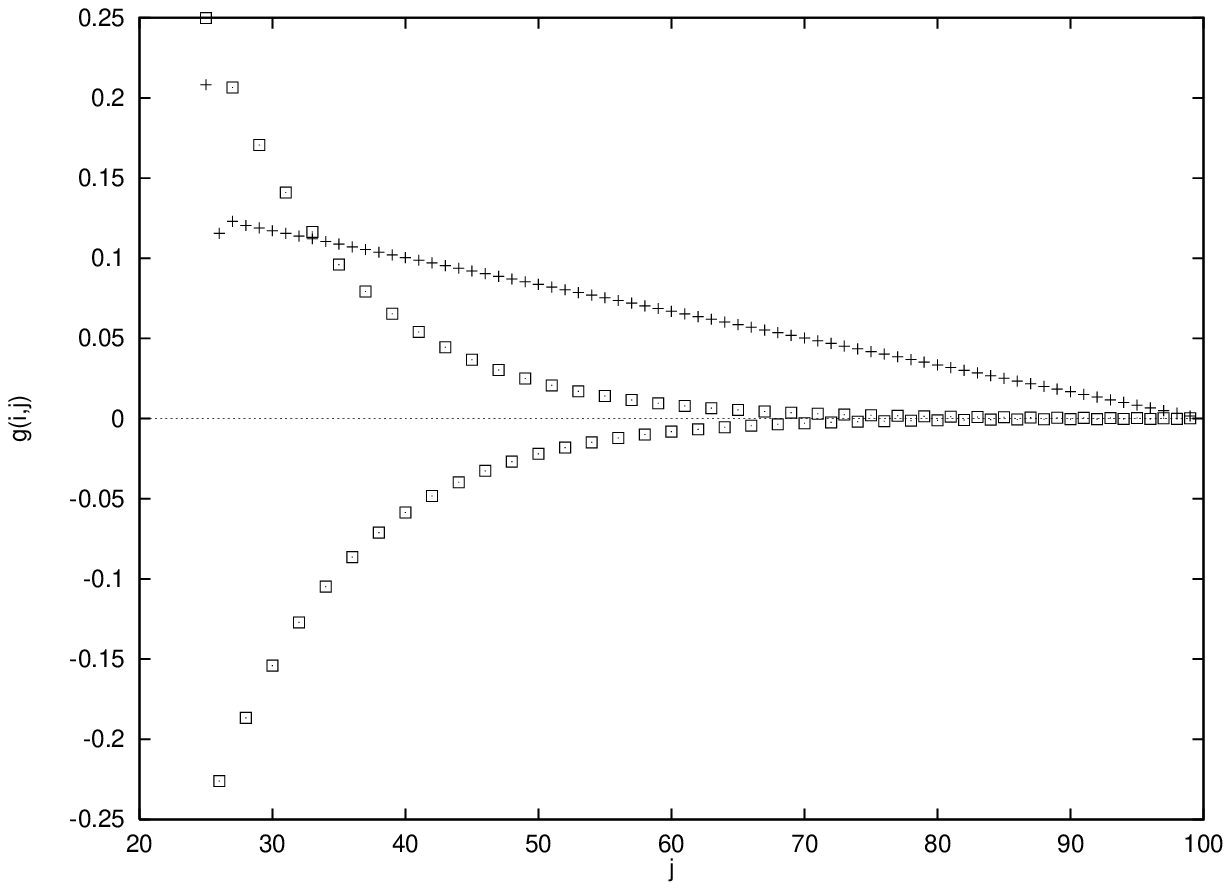}{11cm}{The exact truncated correlation function 
$g(i,j)$ in the case $p=1$,
for $i=25$ versus $j$. The system size is $N=100$. The oscillating
curve (squares) was obtained for $\alpha=\beta=0.9$, the other curve
(crosses) for $\alpha=\beta=0.1$.}{labl}
\begin{figure}[t]
\centerline{\psfig{figure=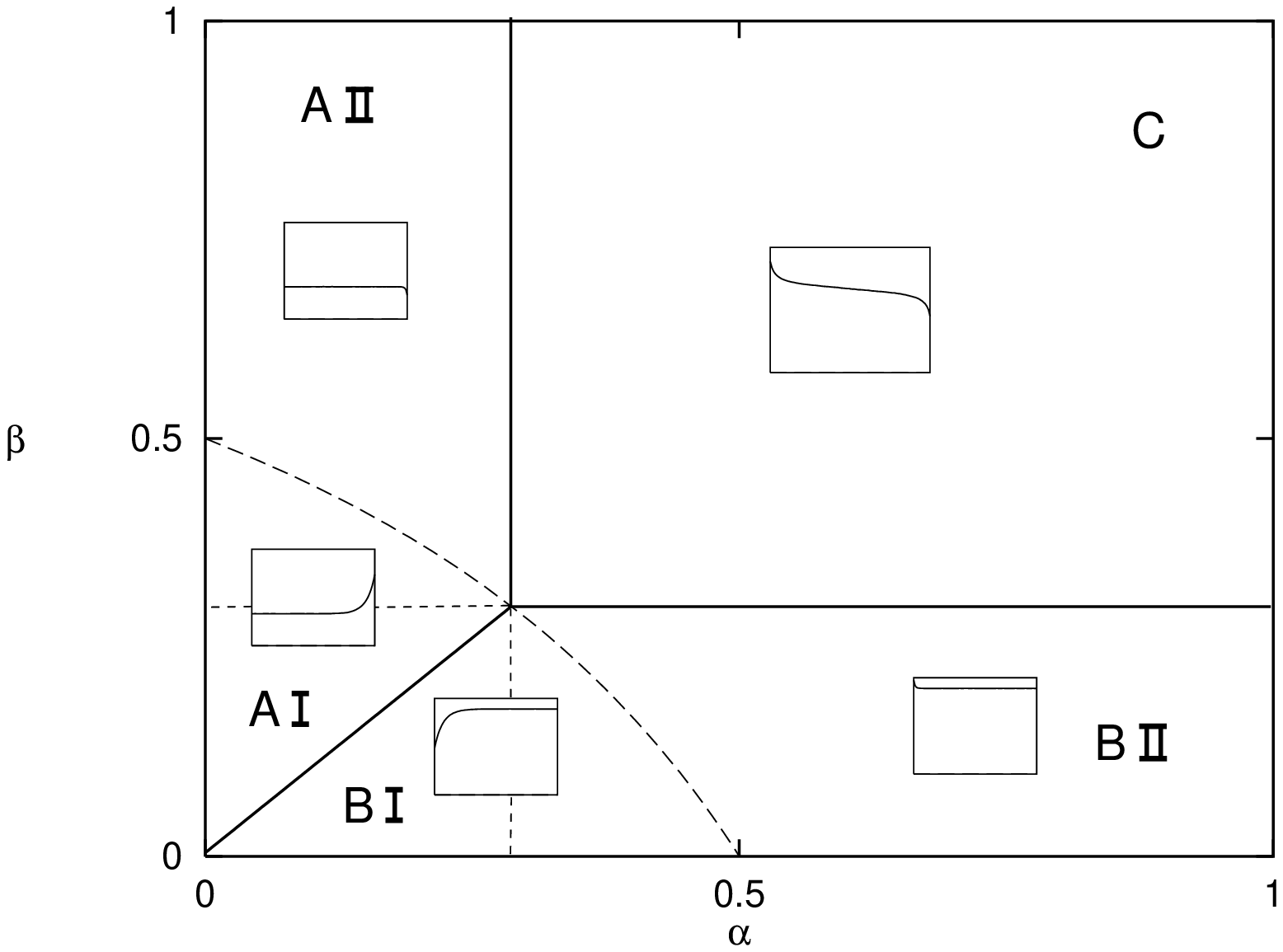,bbllx=10pt,bblly=10pt,
bburx=570pt,bbury=420pt,height=12cm}}
\caption{\small\it (taken from \cite{RSSS}): Phase diagram for the ASEP with 
parallel (synchronous) update 
for $p=0.5$. C is the maximum current
phase, A and B are the low and high density phase, respectively.
The straight dashed lines are the boundaries between 
phase A$\,$I and A$\,$II (B$\,$I and B$\,$II). The curved dashed line
is the line given by (\ref{specialline}) and intersects the line
$\alpha=\beta$ at $\alpha=\beta=1-\sqrt{1-p}=q$ (see section
\ref{phasediagram}).
The inserts show typical 
density profiles in the various phases; note that the profile is
qualitatively the same in region A$\,$I (B$\,$I) 
and in the portion of region A$\,$II (B$\,$II)
below the curved dashed line.}
\label{X1}
\end{figure}
\end{document}